\documentclass[11pt]{article}
\usepackage[centertags]{amsmath}
\usepackage{amsfonts}
\usepackage{newlfont}
\usepackage{graphicx}
\usepackage[dvips]{color}

\newtheorem{proposition}{Proposition}[section]

\newtheorem{ex}{Example}[section]

\newtheorem{rem}{Remark}[section]
\newcommand{\E}{\mathbf{E}}

\newcommand{\e}{\mathrm{e}}
\newcommand{\ci}{\mathrm{i}}

\title{Fourier Transform Methods for Regime-Switching Jump-Diffusions
and the Pricing of \emph{Forward Starting} Options}
\author{A. Ramponi}

\date{Department of Financial Economics and Quantitative Methods \\ School of Economics \\ University of Roma - Tor
Vergata \\ via Columbia, 2, 00133 - Roma, Italy \\ e-mail:
ramponi@economia.uniroma2.it}

\begin{document}

\maketitle

\begin{abstract}
In this paper we consider a jump-diffusion dynamic whose
parameters are driven by a continuous time and stationary Markov
Chain on a finite state space as a model for the underlying of
European contingent claims. For this class of processes we firstly
outline the Fourier transform method both in log-price and
log-strike to efficiently calculate the value of various types of
options and as a concrete example of application, we present some
numerical results within a two-state regime switching version of
the Merton jump-diffusion model. Then we develop a closed-form
solution to the problem of pricing a Forward Starting Option and
use this result to approximate the value of such a derivative in a
general stochastic volatility framework.

\medskip



\end{abstract}

\section{Introduction}

Since the paper by Naik (1993), the use of  continuous time
regime-switching processes to model asset price dynamics
stimulated an increasing interest in the context of option
pricing. The empirical evidence of a regime switching behavior of
some economic time series was pointed out by Hamilton (1989,
1990), who suggested the use of an underlying Markov chain
switching between \emph{regimes} to account for some peculiarities
in observed data. The ability of these econometric models to
capture specific features such as volatility clustering and
structural breaks is widely recognized (see e.g. Timmermann
(2000)). Consequently, they can be considered as an appealing
class of models also in the framework of derivative pricing. In
the last decades there has been a considerable progress in the
pricing exercise for plain vanilla European or American style
options: see e.g. Di Masi et al. (1994), Bollen (1998), Guo
(2001), Hardy (2001), Duan et al. (2002), Buffington and Elliott
(2002), Konikov and Madan (2002), Guo and Zhang (2004), Chourdakis
(2004,2007), Edwards (2005), Liu et al. (2006), Yao et al. (2006),
Jobert and Rogers (2006), Elliott and Osakwe (2006), Jiang and
Pistorius (2008), Boyarchenko and Levendorskii (2009), Khaliq and
Liu (2009), Di Graziano and Rogers (2009), Ramponi (2009), Liu
(2010). Comparatively few results are available for exotic
options: see Boyle and Draviam (2007) and Elliott et al. (2007).
Such results typically differ in the model considered (switching
diffusions or more general L\'{e}vy processes), in the technique
for solving the pricing problem (direct evaluation of expectations
with respect to the probability density of the underlying,
numerical solution of the associated PDE, recombining trees,
Fourier transform methods) or in the type of financial product to
price.

In this paper we consider a quite general underlying dynamic which
can be seen as a switching L$\acute{\textrm{e}}$vy process of the
I type, or finite activity L$\acute{\textrm{e}}$vy process (see
e.g. Cont and Tankov (2004)). In particular, on a filtered
probability space $(\Omega, \{\mathcal{F}_t \}, \mathcal{F},
\mathcal{P})$ the dynamic is of the form
\begin{equation}
    S(t) = s_0 \e^{X(t)}
\end{equation}
where $X(t)$ is specified as a jump-diffusion whose parameters are
driven by $\alpha(t)$, a continuous time and stationary Markov
Chain on the state space $\mathcal{S} = \{1,2,\ldots,M \}$. This
model provides an example of non-affine and non-L\'{e}vy process
for which we are able to calculate the characteristic function
(see Prop. \ref{propCFX}) and therefore the pricing problem for
European style options is efficiently faced through Fourier
transform techniques. Such techniques, originated by the works of
Heston (1993) and Carr and Madan (1999), are based on the
representation of the value of the option in a proper Fourier
space and have been successfully applied to a variety of pricing
problems in the last years. Among the various contributions to
this theory, see Bakshi and Madan (2000), Raible (2000), Lewis
(2002), Lee (2004), Hubalek et al. (2006), Biagini et al. (2008),
and more recently Cherubini et al. (2009), Dufresne et al. (2009),
Hurd and Zhou (2010), Eberlein et al. (2010). Following this
approach we can price various types of European options under the
regime-switching dynamic by using the Fourier transform method
both in the log-price space and in the log-strike space,
consequently taking advantages from the powerful Fast Fourier
Transform (FFT) computational tool. The case for a switching pure
jump process has been considered in Elliott and Osakwe (2006).

As an application we consider the problem of valuing a Forward
Starting option (FSO) for which an almost (i.e. up to numerical
integration) closed-form solution is obtained in term of an
integral transform. A similar technique was used in Kruse and
N\"{o}gel (2005) to price a FSO in the Heston stochastic
volatility model. These options are well-known exotic derivatives
(see e.g. Hull (2009)) characterized by the payoff
\begin{equation}
\Pi_T(S(T),\kappa) = S(T) - \kappa S(t^*)
\end{equation}
where $t^* \in (0,T)$ is the \emph{determination time} and $\kappa
\in (0,1)$ is a given percentage. They are the building blocks of
the so-called cliquet options.  As it will be shown, our formula
is very simple, being a finite mixture of call prices evaluated at
the determination time under each regime, weighted by the
stationary probability of the chain. Furthermore, in Chourdakis
(2004) a procedure to approximate the value of an European option
in a model with stochastic volatility and jumps was proposed by
building a continuous-time Markov chain which "mimics" the
volatility process. The approximating dynamics turns out to be a
regime-switching jump-diffusion model. By using such an
approximation, a pricing algorithm for FSO in a general stochastic
volatility model can be designed based on our mixture
representation.

The paper is organized as follows. In Section 2 the dynamic model
for the underlying is presented with a scheme for its numerical
simulation and a useful representation through the sojourn times
of the underlying Markov chain is introduced. In Section 3 the
Fourier transform method both in log-price and log-strike is
considered and formulas for the price of an European call option
are explicitly derived. A numerical example of calibration on real
data for a two-state regime switching jump diffusion model with
gaussian jumps is reported. Finally, in Section 4 the price of a
Forward starting option is obtained by using the Fourier transform
representation and an algorithm to get approximate prices in a
general stochastic volatility model is outlined.

\section{The model}

Let us consider on a filtered probability space $(\Omega,
\{\mathcal{F}_t \}, \mathcal{F}, \mathcal{P})$, the asset price
dynamic of the form
\begin{equation}
    S(t) = s_0 \e^{X(t)}
\end{equation}
where $X(t)$ is specified as follows.

Let $\alpha(t)$ be a continuous time, homogeneous and stationary
Markov Chain on the state space $\mathcal{S} = \{1,2,\ldots,M \}$
with a generator $Q \in \mathbb{R}^{M\times M}$; furthemore $\xi :
\mathcal{S} \rightarrow \mathbb{R}$, $\sigma : \mathcal{S}
\rightarrow \mathbb{R}$ and $\gamma : E \times \mathcal{S}
\rightarrow \mathbb{R}$ are given functions, $(E,\mathcal{E})$
being a measurable mark space. In a given interval $0 \leq t \leq
T$, we consider the following dynamic
\begin{eqnarray*}
d X(t) & = & \xi(\alpha(t)) dt + \sigma(\alpha(t)) d
W(t) + dJ(t), \ \ X(0)=0, \\
J(t) & = & \int_0^t \int_E \gamma(y,\alpha(s-))p^{\alpha}(dy,ds)
\end{eqnarray*}
where $p^{\alpha}(dy,ds)$ is a marked point process (Runggaldier
(2003)) characterized by the intensity
$$
\lambda(\alpha,dy) = \lambda(\alpha) m(\alpha,dy).
$$
Here $\lambda(\cdot)$ represents the (regime-switching) intensity
of the Poisson process $N_t(E)$, while $m(\cdot,dy)$ are a set of
probability measures on $(E,\mathcal{E})$, one for each state
(regime) $i \in \mathcal{S}$ of the chain. The function
$\gamma(y,\alpha)$ represents the jump amplitude relative to the
mark $y$ in regime $\alpha$. Throughout the paper we assume that
the processes $\alpha(\cdot)$ and $W(\cdot)$ are independent and
that $W(\cdot)$ and $p^{\alpha}(dy,dt)$ are conditionally
independent given $\alpha(t)$. We denote $\mathcal{F}^{\alpha}_t =
\sigma \{\alpha(s): 0 \leq s \leq t \}$ the $\sigma$-algebra
generated by the Markov chain. Furthermore, we assume that
$\E[\e^{\gamma(Y(\alpha),\alpha)}] \equiv \int_E
\e^{\gamma(y,\alpha)} m(\alpha,dy)$ is finite for each regime
$\alpha$, where $Y(\alpha)$ is the random variable associated to
the measure $m(\alpha,dy)$. We also define the compensated point
process $q^{\alpha}(dy,dt) = p^{\alpha}(dy,dt)-\lambda(\alpha(t-))
m(\alpha(t-),dy) dt$ in such a way
$$
\int_0^t \int_E H(y,\alpha(s-))q^{\alpha}(dy,ds)
$$
is a martingale in $t$ for each predictable process $H$ satisfying
appropriate integrability conditions. In particular the jump
process
\begin{eqnarray*} J(t) & = & \int_0^t \int_E
\gamma(y,\alpha(s-))q^{\alpha}(dy,ds) + \int_0^t \int_E
\gamma(y,\alpha(s)) \lambda(\alpha(s)) m(\alpha(s),dy) ds \\
 & = & \tilde J(t) + \int_0^t
\lambda(\alpha(s)) \E[\gamma(Y(\alpha(s)),\alpha(s))] ds, \ \ \
Y(\alpha(s)) \sim m(\alpha(s),dy)
\end{eqnarray*}
is the sum of a martingale and an absolutely continuous process,
whenever $\gamma$ satisfies the proper conditions.

\bigskip

A sample path of this process is generated as follows (see Fig.1):
\begin{enumerate}
\item generate a path of the Markov chain, i.e. a set
of switching times $\tau_0=0, \tau_1, \ldots, \tau_L,
\tau_{L+1}=T$ and the corresponding states $\alpha(t) = \alpha_k
\in \mathcal{S}, \tau_k \leq t < \tau_{k+1}, k=0,\ldots,L$;
\item generate the jump times $\upsilon_{k_j}$ of the Poisson process in each interval
$[\tau_k, \tau_{k+1})$ according to the intensity
$\lambda(\alpha_k)$ and let $N_k$ be the number of jumps;
\item for any $k=0, \ldots, L$ generate $N_k$ i.i.d. samples
$Y(\alpha(\upsilon_{k_j})), j=1,\ldots,N_k$ distributed according
to the probability $m(\alpha_k,dy)$;
\item on a given time grid $t_0, \ldots, t_n$ of $[0, T]$ built as
the superposition of a deterministic grid and the jump times
$\upsilon_j$, let $X(t_0)=0$ and
\begin{eqnarray}
X(t_{i+1}-) & = & X(t_i) + \xi(\alpha(t_i)) (t_{i+1}-t_i)
+ \sigma(\alpha(t_i)) (W(t_{i+1}-W(t_i)), \\
X(t_{i+1}) & = & X(t_{i+1}-) + \int_E \gamma(y,\alpha(t_{i+1}))
p(dy,t_{i+1}).
\end{eqnarray}
If $t_{i+1}$ is actually a point of the Poisson random measure,
the magnitude of the jump is sampled, that is
$$
\int_E \gamma(y,\alpha(t_{i+1})p(dy,t_{i+1}) =
\gamma(Y(\alpha(t_{i+1})),\alpha(t_{i+1})),
$$
otherwise the jump term is zero.
\end{enumerate}

\begin{figure}[t]
\begin{center}
\includegraphics[width=12cm,height=10cm]{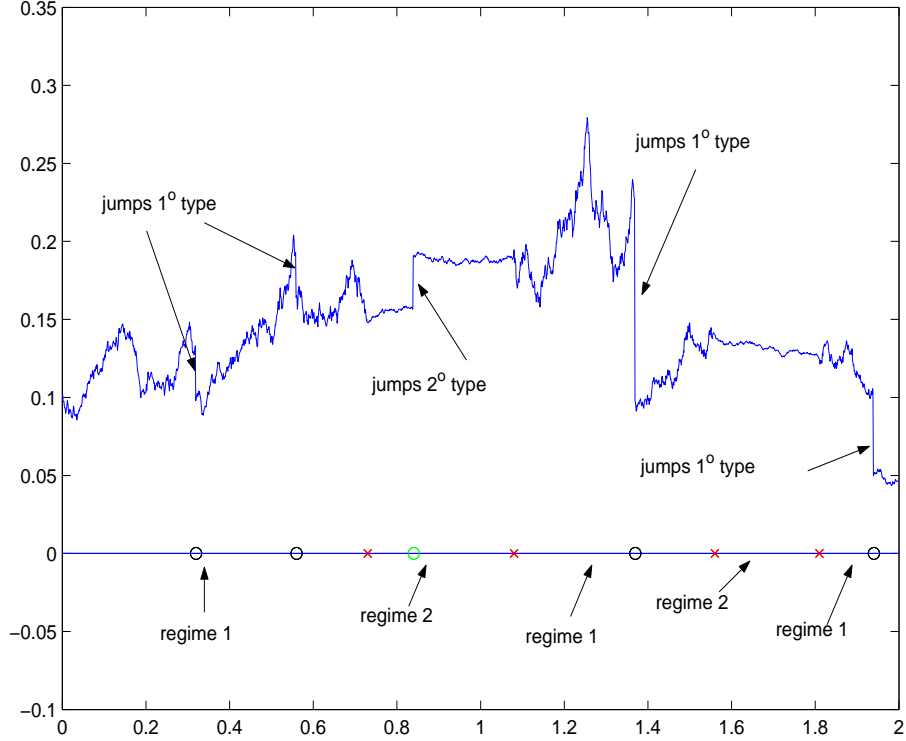}
\caption{\small A sample path of the RSJD - Example
\ref{ex_2rsjdmerton}. Red cross are the switching times, circles
represent the jump times for the two regimes.} \label{fig1}
\end{center}
\end{figure}

\bigskip

In view of our pricing application, from now on we assume to
specify our model in a probability space where the "discounted"
asset price $\tilde S(t) = S(t) \e^{-\int_0^t \mu(\alpha(s))ds}$
is a martingale. In particular, we keep the function $\mu :
\mathcal{S} \rightarrow \mathbb{R}$ unspecified in order to cope
with slightly different types of contracts: for example, given the
risk-free rate $r$, we can set $\mu(\alpha) = \mu$ with $\mu=
r-q$, $q$ being the (continuous) dividend rate, $\mu = r-r_f$,
$r_f$ being the foreign risk-free rate or more generally
$\mu(\alpha) = r(\alpha)- q(\alpha)$ if rates and dividend are
regime-switching too.

We therefore consider the following model
\begin{eqnarray} \label{rsxdyn}
\nonumber X(t) & = & \int_0^t \left( \mu(\alpha(s)) - \frac{1}{2}
\sigma^2(\alpha(s)) - \lambda(\alpha(s))
\kappa(\alpha(s))\right )ds + \int_0^t \sigma(\alpha(s)) dW(s) \\
\label{rsjd1} & + & \int_0^t \int_E
\gamma(y,\alpha(s^-))p^{\alpha}(dy,ds),
\end{eqnarray}
where
$$
\kappa(\alpha) = \E[(\e^{\gamma(Y(\alpha),\alpha)}-1)], \ \ \alpha
\in \mathcal{S}.
$$

An application of the generalized Ito's Formula gives
\begin{eqnarray*}
d \tilde S(t) & = & \tilde S(t-) \left( - \lambda(\alpha(t))
\kappa(\alpha(t)) dt + \sigma(\alpha(t)) dW(t) + \int_E
(\e^{\gamma(y,\alpha(t-))}-1)
p^{\alpha}(dy,dt) \right) \\
& = & \tilde S(t-) \left(\sigma(\alpha(t)) dW(t) + \int_E
(\e^{\gamma(y,\alpha(t-))}-1)q^{\alpha}(dy,dt) \right),
\end{eqnarray*}
where $q^{\alpha}(dy,dt) = p^{\alpha}(dy,dt)- \lambda(\alpha(t-))
\kappa(\alpha(t-))dt$ is the compensated process. Hence, $\tilde
S(t)$ is a martingale. The corresponding jump-diffusion SDE for
the asset price is therefore
\begin{equation}\label{rsjdm_sde}
\frac{d S(t)}{S(t-)} = (\mu(\alpha(t)) - \lambda(\alpha(t))
\kappa(\alpha(t)) ) dt + \sigma(\alpha(t)) dW(t) + \int_E
(\e^{\gamma(y,\alpha(t-))}-1)p^{\alpha}(dy,dt), \ \ \ S(0)=s_0.
\end{equation}

{\ex \label{ex_2rsjdmerton} As a working example we consider a
two-state regime switching version of the Merton jump-diffusion
model. This is defined by taking $\gamma(y,\alpha)=y$ and two
kinds of normal jumps, i.e. $Y(i) \sim \mathcal{N}(a_i,b_i)$ from
which $\kappa(i)= \E[(\e^{Y(i)}-1)] = \e^{a_i + b_i^2/2}-1$,
$i=1,2$. The two state Markov chain $\alpha(t) \in \mathcal{S} =
\{1,2 \}$ has
generator $Q = \left(%
\begin{array}{cc}
  -q_1 & q_1 \\
  q_2 & -q_2 \\
\end{array}%
\right)$. Let $\sigma_i, \lambda_i > 0$ and $\mu_i, i=1,2$ be
given parameters: the regime switching jump-diffusion Merton model
is defined as
\begin{eqnarray*} \label{merton1}
dX(t) & = & [\mu(\alpha(t)) - \frac{1}{2} \sigma^2(\alpha(t) -
\lambda(\alpha(t)) \kappa(\alpha(t))]dt + \sigma(\alpha(t)) dW(t)
+ dJ(t) \\
J(t) & = & \int_0^t \int_E  y p^{\alpha}(dy,ds), \ \ \
\lambda(t,\alpha(t),dy) = \lambda(\alpha(t)) \phi_{\alpha(t)}(y)
dy
\end{eqnarray*}
where $\lambda(\alpha(t)) \in \{\lambda_1,\lambda_2 \}$,
$\sigma(\alpha(t)) \in \{\sigma_1,\sigma_2 \}$, $\mu(\alpha(t))
\in \{\mu_1,\mu_2 \}$ and $\lambda(t,\alpha(t),dy)$ is the
intensity process of the Poisson jump component, $\phi_{i}(y)$
being the density of a normal distribution $\mathcal{N}(a_i,
b_i)$, $i=1,2$. $\Box$}

\bigskip

Next Proposition gives a useful representation for $X(T)$. A
sketch of the proof is reported in the Appendix.

{\proposition \label{Xrepres} Let $T_i = \int_0^T
\mathbb{I}_{\alpha(s)=i} ds$, $i=1,\ldots,M$ be the occupation
times of the Markov chain and let us define $\xi(\alpha)
=\mu(\alpha) - \frac{1}{2} \sigma^2(\alpha) - \lambda(\alpha)
\kappa(\alpha)$ and
$$
\Xi_T(T_1, \ldots,T_M) = \int_0^T \xi(\alpha(s)) ds = \sum_{i=1}^M
\xi(i) T_i.
$$

Then the process $X(T)$ admits the following representation:
\begin{equation} \label{rsjd2}
X(T) = \Xi_T(T_1, \ldots,T_m) + \sum_{i=1}^M \sigma(i) Z(\Delta_i)
+\sum_{i=1}^M \sum_{k=1}^{N(\Delta_i)} Y^{(i)}_k,
\end{equation}
where $N(\Delta_i)$ and $Z(\Delta_i)$ are distributed as Poisson
variables $\mathrm{Poiss}(\lambda_i T_i)$ and as Normal variables
$\mathcal{N}(0,T_i)$, respectively, $i=1,\ldots, M$.}

\bigskip

It is readly seen that by defining $X_{t,T}= X(T)-X(t)$ we have
\begin{equation} \label{rsjd3}
X_{t,T}= \Xi_{t,T}(T^{t,T}_1, \ldots,T^{t,T}_M) + \sum_{i=1}^M
\sigma(i) Z(\Delta_i) +\sum_{i=1}^M \sum_{k=1}^{N(\Delta_i)}
Y^{(i)}_k,
\end{equation}
where $T^{t,T}_j = \int_t^T \mathbb{I}_{\alpha(s)=j} ds$,
$j=1,\ldots, M$, and the random variables $N(\Delta_i)$ and
$Z(\Delta_i)$ have conditional distributions
$\mathrm{Poiss}(\lambda_i (T_i-t))$ and $\mathcal{N}(0,T_i-t)$.
Correspondingly we can write $S(T)=S(t) \exp(X_{t,T})$.

{\rem Notice that for a L$\acute{\textrm{e}}$vy process $X_L(t)$
having characteristic function
$$
\E[\e^{\ci u X_L(t)}] = \exp \left( t (\ci \xi u - \frac{\sigma^2
u^2}{2} + \int_{\mathbb{R}}(e^{\ci u x}-1) \beta(dx)) \right)
$$
the expected value is $ \E[X_L(T)] = T (\xi + \int_{\mathbb{R}} x
\beta(dx))$. For our RS model, it follows from (\ref{Xrepres})
that
$$
\E[X(T)] = \sum_{i=1}^m (\xi(i) + \lambda(i) \mu_{Y^{(i)}})
\E[T_i] = \sum_{i=1}^m (\xi(i) + \lambda(i) \mu_{Y^{(i)}})
\int_0^T \mathcal{P}(\alpha(s)=i | \alpha(0)) ds.
$$

This quantity can be easily evaluated for a two-state MC, since we
have $P_s = \e^{Q s} =
\frac{1}{\mu + \nu}  \left(%
\begin{array}{cc}
  \mu \e^{-s(\mu+\nu)}+\nu & \mu(1-\e^{-s(\mu+\nu)}) \\
  \nu(1-\e^{-s(\mu+\nu)}) & \nu \e^{-s(\mu+\nu)}+\mu \\
\end{array}%
\right)$. This implies that, starting e.g. from $\alpha(0)=1$
\begin{eqnarray*}
\E[X(T)]  & = & (\xi(1) + \lambda(1) \mu_{Y^{(1)}})
(\frac{\mu}{\mu+\nu} \frac{1-\e^{-T(\mu+\nu)}}{\mu+\nu} +
\frac{\nu}{\mu+\nu} T) \\ & + & (\xi(2) + \lambda(2)
\mu_{Y^{(2)}}) (-\frac{\mu}{\mu+\nu}
\frac{1-\e^{-T(\mu+\nu)}}{\mu+\nu} + \frac{\mu}{\mu+\nu} T).
\end{eqnarray*}
}

\section{The transform method}

Our main interest is the efficient numerical evaluation of the
price $\Pi_0$ of an European contingent claim specified by the
payoff function $\Pi(s,K)$, exercised at the future time $T$, $K$
being a trigger parameter. By letting $r(t)$ be the interest rate
process, $B(t)=\exp(\int_0^t r(u) du)$ the usual money market
account and $P(t,T)$ the time-$t$ value of a discount bond
maturing at $T$, arbitrage pricing theory and a
change-of-numeraire technique give the well-known characterization
of prices
$$
\Pi_0 = \E^{\mathcal{P}}[B(T)^{-1} \Pi(S(T),K)] = P(0,T)
\E^{\mathcal{Q}}[\Pi(S(T),K)], \ \ \ P(0,T) =
\E^{\mathcal{P}}[B(T)^{-1}],
$$
where $\mathcal{P}$ is the risk-neutral measure and $\mathcal{Q}$
is known as $T$-forward measure. When interest rates are
deterministic, the two measures are equal. In the following we
assume that our dynamic model is given under the measure
$\mathcal{Q}$. All the expected values will be considered with
respect to this measure.

\bigskip

It is well known that Fourier transform methods can be efficiently
used for the valuation of European style options. Two main
variants have been developed depending on which variable of the
payoff is transformed into the Fourier space. In view of the
structure assumed for the dynamic of the underlying price $S(T)$
and our next applications, we consider instead $\log(S(T)) = X(T)
+ \log(s_0)$ as the state variable and $k = \log(K)$ for the
trigger parameter, in such a way for any payoff $\Pi(s,K) =
\Pi(e^{\log (s)}, e^{\log(K)}) \equiv \Pi(y,k)$. Correspondingly,
we can consider the generalized Fourier transform with respect to
the state variable $y$, $\hat \Pi_k(z) = \int_{\mathbb{R}} \e^{\ci
z y} \Pi(y,k) dy$ (log-price transform), or w.r.t. the trigger
$k$, $\hat \Pi_y(z) = \int_{\mathbb{R}} \e^{\ci z k} \Pi(y,k) dk$
(log-strike transform), $z \in \mathbb{C}$. In general we assume
that these transforms exist in some strip $\mathcal{S}_{\Pi}= \{ z
\in \mathbb{C} : -\infty \leq a < \Im(z) < b \leq +\infty
\}$\footnote{$\Im(z)$ and $\Re(z)$ stand for the imaginary and
real part of a complex number, $z=\Re(z)+\ci \Im(z) \in
\mathbb{C}$.} of the complex plane. Examples of payoffs are
reported in Table (\ref{GFTab}). The first approach was proposed
in this form in Raible (2000) (but the representation of option
prices through inversion of characteristic function appeared for
the first time in Heston (1993)), while the second was introduced
in Carr and Madan (1999).

Formally, Fourier inversion gives
$$
\Pi(y,k) = \left \{ \begin{array}{cc}
                      \frac{1}{2\pi} \int_{\ci \nu - \infty}^{\ci \nu +\infty}
                      \e^{-\ci z y} \hat \Pi_k(z) dz, & \mbox{log-price transform} \\ \\
                      \frac{1}{2\pi} \int_{\ci \nu - \infty}^{\ci \nu +\infty}
                      \e^{-\ci z k} \hat \Pi_y(z) dz, & \mbox{log-strike transform} \\
                    \end{array} \right.
$$
where integrals are considered along the straight line $\Im(z) =
\nu$ in the complex plane. By letting $\varphi_T(z) = \E[\e^{\ci z
X(T)}], \ z \in \mathbb{C}$ be the (generalized) Fourier transform
(or characteristic function) of $X(T)$, we have
$$
\Pi_0 / P(0,T) = \E^{\mathcal{Q}}[\Pi(X(T)+\log(s_0),k)] =
\int_\mathbb{R} \Pi(y,k) \mathcal{Q}_T(dy)
$$
$$
= \left\{ \begin{array}{c}
            \int_\mathbb{R} \frac{1}{2\pi} \int_{\ci \nu - \infty}^{\ci \nu +\infty} \e^{-\ci z y} \hat \Pi_k(z) dz \mathcal{Q}_T(dy)
            \\ \\
            \int_\mathbb{R} \frac{1}{2\pi} \int_{\ci \nu - \infty}^{\ci \nu +\infty} \e^{-\ci z k} \hat \Pi_y(z) dz \mathcal{Q}_T(dy) \\
          \end{array} \right. = \left\{ \begin{array}{c}
            \frac{1}{2\pi} \int_{\ci \nu - \infty}^{\ci \nu +\infty} \hat \Pi_k(z) \int_\mathbb{R}  \e^{-\ci z y}
            \mathcal{Q}_T(dy) dz
            \\ \\
            \frac{1}{2\pi} \int_{\ci \nu - \infty}^{\ci \nu +\infty} \e^{-\ci z k}  \int_\mathbb{R} \hat \Pi_y(z) \mathcal{Q}_T(dy) dz \\
          \end{array} \right.
$$
\bigskip
\begin{equation} \label{transfprices}
= \left\{ \begin{array}{ll}
            \frac{1}{2\pi} \int_{\ci \nu - \infty}^{\ci \nu +\infty}  \e^{-\ci z \log(s_0)} \hat \Pi_k(z) \varphi_T(-z)
            dz, & \mbox{log-price transform} \\ \\
            \frac{1}{2\pi} \int_{\ci \nu - \infty}^{\ci \nu +\infty} \e^{-\ci z k}
            \E^{\mathcal{Q}}[\hat \Pi_{X(T)+\log(s_0)}(z)] dz, & \mbox{log-strike transform}.\\
          \end{array} \right.
\end{equation}

\bigskip

In order to justify the previous equalities, some conditions are
required: existence of the generalized Fourier transform $\hat
\Pi$, integrability along the contour $\Im(z) = \nu$ in some strip
$\mathcal{S}_{\Pi}$ in order to guarantee the Inversion Theorem
and existence of the expectation $\E^{\mathcal{Q}}[e^{\nu X(T)}]$
(see Lee (2004) for log-strike transform, Lewis (2002) or the
recent Eberlein et al. (2009) for log-price transform). Notice
that the use of generalized Fourier transform permits to exploit
contour variations by means of the residue theorem, as it will be
seen in next paragraphs.

Due to the exponential structure of the GFT of typical payoffs
(see Table (\ref{GFTab})), also for the log-strike transform it is
required the calculation of $\varphi_T(z)$ appearing through the
expectation $\E^{\mathcal{Q}}[\hat \Pi_{X(T)+\log(s_0)}(z)]$. Next
Proposition gives the GFT of our process. Similar results are
available (see Chourdakis (2004)) where a particular structure of
the generator $Q$ is considered: for completeness, we report the
proof in the Appendix.

\begin{table}[t]
\begin{center}
\begin{tabular}{|c|c|c||c|c|}
  \hline
  Payoff & GFT in log-price & Strip of regularity & GFT in log-strike & Strip of regularity \\ \hline
  $(\e^y-\e^k)^+$ & $\frac{\e^{k(\ci z +1)}}{\ci z-z^2}$ & $\Im(z) > 1$
  & $\frac{\e^{y(\ci z +1)}}{\ci z-z^2}$ &  $\Im(z) < 0$ \\ \hline
  $(\e^k-\e^y)^+$ & $\frac{\e^{k(\ci z +1)}}{\ci z-z^2}$ & $\Im(z) < 0$
  & $\frac{\e^{y(\ci z +1)}}{\ci z-z^2}$ & $\Im(z) > 1$\\ \hline
  $\e^{a y} \mathbb{I}_{by>\kappa}$ & $- \frac{\e^{(a+\ci z) \kappa/b}}{a+\ci z}$  & $\Im(z) > a$
  & $\frac{\e^{(a+\ci z b) y}}{\ci z}$ & $\Im(z) > 0$ \\
  \hline
  $\min (\e^y, \e^k)$ & $\frac{\e^{k(\ci z +1)}}{z^2-\ci z}$ & $0 < \Im(z) <
  1$ & $\frac{\e^{y(\ci z +1)}}{z^2-\ci z}$ & $0 < \Im(z) <  1$ \\
  \hline
\end{tabular}
\caption{Generalized Fourier transforms of typical payoffs.}
\label{GFTab}
\end{center}
\end{table}

{\proposition \label{propCFX} Let $\phi_j(z)=\E[\e^{\ci z
\gamma(Y(j),j)}]$ be the generalized Fourier transform of the jump
magnitude. Then, by letting
\begin{equation}\label{theta1}
\vartheta_j(z) = z \xi(j)+\frac{1}{2} \ci z^2 \sigma^2(j) - \ci
\lambda(j)(\phi_i(z)-1)
\end{equation}
and $\tilde \vartheta_i(z) = \vartheta_j(z) - \vartheta_M(z)$, we
have
\begin{equation} \label{charfun2}
\begin{array}{lll}
\varphi_T(z) & = & \e^{\ci \vartheta_M(z) T} \left(\mathbf{1}'
\cdot \e^{(Q' + \ci \ \mathrm{diag}(\tilde \vartheta_1(z), \ldots,
\tilde \vartheta_{M-1}(z),0))T} \cdot \mathbb{I}(0) \right) \\ \\
 & = & \mathbf{1}' \cdot \e^{(Q' + \ci \
\mathrm{diag}(\vartheta_1(z), \ldots, \vartheta_{M}(z)))T} \cdot
\mathbb{I}(0), \end{array}
\end{equation}
where $\mathbf{1} = (1,\ldots,1)' \in \mathbb{R}^{M \times 1}$,
$\mathbb{I}(0) = (\mathbb{I}_{\alpha(0)=1}, \ldots,
\mathbb{I}_{\alpha(0)=M})'\in \mathbb{R}^{M \times 1}$ and $Q'$ is
the transpose of $Q$. }

{\rem Notice that $\varphi_T(0) = 1$ and $\varphi_T(-\ci) =\E
\left[ \e^{\sum_{j=1}^M \mu(j) T_j} \right]$. Furthermore, if
$\mu(\alpha) \equiv \mu$, then $\varphi_T(-\ci) = \e^{\mu T}$
since $\sum_{i=1}^m T_i = T$. \label{remark1} }

\bigskip

More generally, we get from (\ref{rsjd3}) and (\ref{charfun2})
\begin{equation} \label{charfun3}
\begin{array}{lll}
\varphi_{t,T}(z) & = & \E_t[\e^{\ci z X_{t,T}}] =
\left(\mathbf{1}' \cdot \e^{(Q' + \ci \
\mathrm{diag}(\vartheta_1(z), \ldots,
\vartheta_{M}(z)))(T-t)} \cdot \mathbb{I}(t) \right) \\ \\
 & = & \sum_{j=1}^M \mathbb{I}_{\alpha(t)=j} \mathrm{q}_j^{t,T}(z), \ \
\ \ \ \mathrm{q}_j^{t,T}(z) = \sum_{k=1}^M \left(\e^{(Q' + \ci \
\mathrm{diag}(\vartheta_1(z), \ldots,
\vartheta_{M}(z)))(T-t)}\right)_{kj},
\end{array}
\end{equation}
for any $t \in [0,T)$, $\E_t$ being the conditional expectation up
to time $t$. Notice that the characteristic function of $X(T)$ and
$X_{t,T}$ depends on the state of the Markov chain $\alpha(0)$ and
$\alpha(t)$, respectively.

\medskip

The conditions for applying the transform method both in log-price
and log-strike depend on the properties of the GTF of $X(T)$,
which in turn depend on that of $\phi_j(z)$ through Proposition
\ref{propCFX}. In general, these functions are well defined (and
analytic) in some strips of the complex plane
$$
\mathcal{S}_j= \{ z \in \mathbb{C} : \E^{\mathcal{Q}}[e^{\Im(z)
 \gamma(Y(j),j)}]<  \infty \}, \ \ j=1,\ldots,M.
$$

Let us define the matrix $A(z) =Q' + \ci \
\mathrm{diag}(\vartheta_1(z), \ldots, \vartheta_{M}(z))$: clearly
the elements of $A(z)^n, n=1,2,\ldots$ are polynomials in the
$\vartheta_j(z)$'s and therefore these are well defined in the
intersection of the $\mathcal{S}_j$, $j=1,\dots,M$. From the
properties of the matrix exponential function
$e^{A(z)}=\sum_{n=1}^{+\infty} \frac{A(z)^n}{n!}$ and since the
GTF of $X(T)$ is a linear combination of its elements, it
immediately follows that (\ref{charfun2}) and (\ref{charfun3}) are
well defined in $\bigcap_{j=1}^M \mathcal{S}_j$ and consequently
the transform methods can be applied, provided $\bigcap_{j=1}^M
\mathcal{S}_j \neq \emptyset$ and the payoffs satisfy the proper
conditions.

{\rem \label{remark2} If we set $\mu(i)=\mu, \sigma(i)=\sigma$,
$\lambda(i)=\lambda$ and $\phi_i(z) =\phi(z)$ we have that $\tilde
\vartheta_i(z)=0$, $i=1,\ldots, m-1$, and the term $\exp(Q'T)$ is
the transpose of the transition semi-group of the Markov chain.
Under these choices we are implicitly assuming a unique regime and
eq. (\ref{charfun3}) becomes the well-known characteristic
function of the (single-regime) jump-diffusion dynamic
(\ref{rsxdyn}), $\varphi_T(z) = \exp( z \xi+\frac{1}{2} \ci z^2
\sigma^2 - \ci \lambda(\phi(z)-1))$. This is because $\mathbf{1}'
\cdot \e^{(Q' + \ci \ \mathrm{diag}(\tilde \vartheta_1(z), \ldots,
\tilde \vartheta_{M-1}(z),0))(T-t)} \cdot \mathbb{I}(t) =
\mathbf{1}' \cdot \e^{Q'(T-t)} \cdot \mathbb{I}(t)=\sum_{i=1}^M
\mathbb{I}_{\alpha(t)=i} = 1$. Hence, with simple linear
constraints on the full parameter set of our dynamic
(\ref{rsxdyn}) we can recover several models:
\begin{enumerate}
\item Black \& Scholes model (BS): $\mu_i=r$, $\sigma_i=\sigma>0$,
$\lambda_i=0$ (we consequently set to zero the jump variables
$Y(i)$), $i=1,\ldots,M$;
\item Black \& Scholes with regime switching model (RSBS): $\mu_i \in \mathbb{R}$, $\sigma_i>0$,
$q_{ij}>0, i \neq j$, $\lambda_i=0$ ($Y(i)\equiv 0$),
$i=1,\ldots,M$;
\item Merton jump-diffusion model (JDM): $\mu_i=r$, $\sigma_i=\sigma>0$,
$\lambda_i=\lambda >0$ and the parameters of the jump variables
$Y_i\equiv Y$, $i=1,\ldots,M$;
\item Merton jump-diffusion model with regime switching (RSJDM): $\mu_i \in \mathbb{R}$,
$\sigma_i>0$, $q_{ij}>0, i \neq j$, $\lambda_i >0$ and the
parameters of the jump variables $Y(i)$ for each regime,
$i=1,\ldots,M$.
\end{enumerate}
}

\bigskip

From a computational viewpoint, for a fixed complex $z$ the
calculation of $\varphi_T(z)$ requires the following steps:
\begin{enumerate}
\item calculate $\vartheta_j(z), j=1,\ldots, M$ (eq. (\ref{theta1}));
\item form the matrix $A(z)=Q'+ \ci \ \mathrm{diag}(\vartheta_1(z), \ldots, \vartheta_{M}(z))$;
\item calculate the matrix exponential $\Phi(z) = \exp(A(z) \ T)$;
\item for each starting state of the chain $\alpha(0)=j, j=1,\ldots,M$, calculate
$\mathrm{q}_j^{T}(z) = \sum_{k=1}^M \Phi_{kj}(z)$.
\end{enumerate}

For $M > 2$ the cumbersome task is the calculation of $\Phi$ for
which efficient numerical techniques are available (see Higham
(2009)).

\paragraph{The case M=2.} In this case it is possible to give a closed form
solution to the matrix exponential, therefore obtaining an
easy-to-implement formula for the characteristic function. The
following result can be proved either by solving a couple of ODE,
as in Buffington and Elliott (2002) - Appendix 1, or through a
Laplace Transform - based technique, as in Liu et al. (2006).

{\proposition \label{twocharfun} Let $y_{1,2}$ be the solutions of
the quadratic equation $y^2+(q_1+q_2-\ci \theta) y - \ci \theta
q_2 = 0$ and
$$
\begin{array}{lll}
\mathrm{q}_1^{t,T}(\theta) & = & \frac{1}{y_1-y_2} \left( \e^{y_1
(T-t)}(y_1+q_1+q_2)-\e^{y_2 (T-t)} (y_2+q_1+q_2)\right) \\ \\
\mathrm{q}_2^{t,T}(\theta) & = & \frac{1}{y_1-y_2} \left(\e^{y_1
(T-t)}(y_1+q_1+q_2-\ci \theta)-\e^{y_2 (T-t)} (y_2+q_1+q_2-\ci
\theta) \right).
\end{array}
$$

Then
\begin{eqnarray*}
\E_t[\e^{\ci \theta T_1}] = \mathbb{I}_{\alpha(t)=1}
\mathrm{q}_1^{t,T}(\theta)  +  \mathbb{I}_{\alpha(t)=2}
\mathrm{q}_2^{t,T}(\theta).
\end{eqnarray*}
$ \Box$ }

\bigskip

It is easy to prove that the functions $\mathrm{q}_1^{t,T}$ and
$\mathrm{q}_2^{t,T}$ are invariant by changing the order of the
roots $y_1$ and $y_2$. The characteristic function follows from
the proof of Prop. \ref{propCFX} (see (\ref{charfun4})):
\begin{equation} \label{gft2}
\varphi_{t,T}(z) = \e^{\ci \vartheta_2(z) (T-t)} \left(
\mathbb{I}_{\alpha(t)=1} \mathrm{q}_1^{t,T}(\theta(z)) +
\mathbb{I}_{\alpha(t)=2} \mathrm{q}_2^{t,T}(\theta(z))  \right).
\end{equation}

{\ex In our regime switching version of the Merton model we have
$\phi_i(z) = \e^{\ci z a_i - \frac{1}{2} z^2 b^2_i}$, $i=1,\ldots,
M.$ Then, from (\ref{theta1})
\begin{equation}
\vartheta_i(z) = z \xi_i+\frac{1}{2} \ci z^2 \sigma^2_i - \ci
\lambda_i(\e^{\ci z a_i - \frac{1}{2} z^2 b^2_i}-1), \ \
i=1,\ldots,M.
\end{equation}
It follows that in such a case the characteristic function
$\varphi_{T}(z)$ is well defined for all $z \in \mathbb{C}$. In
the two state model the GFT is easily obtained from (\ref{gft2}).
}

\bigskip


\bigskip

Some examples of payoff transforms for the typical claims are
recalled in Table \ref{GFTab}. In view of our next applications,
we show in some details how to get the price of call and put
options both in log-price and in log-strike transform. Pricing
formulas for the other payoffs are reported in Table
\ref{TABVLOGP}.

\paragraph{Call/Put value in $\log$-price transform.} From
formula (\ref{transfprices}) and Table \ref{GFTab} we get for the
call option
\begin{equation} \label{callprice_logprice}
C_0 = \frac{P(0,T)}{2 \pi} \int_{\ci \nu - \infty}^{\ci \nu +
\infty} \e^{-\ci z \log(s_0)} \varphi_T(-z) \frac{\e^{k(\ci z
+1)}}{\ci z - z^2} dz, \ \ \nu
> 1,
\end{equation}
$$
= \frac{P(0,T)}{2 \pi} \ \e^{\nu \log(s_0) + k (1-\nu)}
\int_{-\infty}^{+\infty} \e^{-\ci u (\log(s_0)-k)}
\frac{\varphi_T(-u -\ci \nu)}{\nu^2-\nu -u^2 + \ci u(1-2\nu)} du
$$
$$
=\frac{P(0,T)}{2 \pi} \ s_0^{\nu} K^{1-\nu}
\int_{-\infty}^{+\infty} \e^{-\ci u \log(s_0/K)}
\frac{\varphi_T(-u -\ci \nu)}{\nu^2-\nu -u^2 + \ci u(1-2\nu)}  du,
$$
provided the characteristic function evaluated in the integral
(\ref{callprice_logprice}) is well defined for $z \in \mathbb{C}$
such that $\Im(z) >1$. By switching from $\Im(z) > 1$ to $\Im(z)
<0$ we get the value for the put option: notice that the put-call
parity relation is recovered by moving the integration contour. As
a matter of fact, alternative formulas can be derived by using
residue calculus (see e.g. Lewis(2002)), under the proper
conditions for $\varphi_T(z)$. The GFT of this payoff has two
simple poles at $z = 0$ and $z =\ci$ with residue $-\frac{K \ci}{2
\pi}$ and $\frac{s_0 \varphi_T(-\ci) \ci}{2 \pi}$, respectively:
by moving the integration contour and since the integral must be
real, we obtain the following general formula in which we stress
the dependence on $s_0$, $\alpha_0$ and $K$:
\begin{equation}\label{callprice1}
C_0(s_0,\alpha_0,K)  =  P(0,T)\left(R_\nu + \frac{1}{2 \pi}
\int_{\ci \nu - \infty}^{\ci \nu + \infty} \e^{-\ci z \log(s_0)}
\varphi_T(-z) \frac{\e^{k(\ci z +1)}}{\ci z - z^2}  dz \right)
\end{equation}
$$
=  P(0,T)\left(R_\nu + \frac{1}{\pi} s_0^{\nu} K^{1-\nu}
\int_0^{+\infty} \Re\left[\e^{-\ci u \log(s_0/K)}
\frac{\varphi_T(-u -\ci \nu)}{\nu^2-\nu -u^2 + \ci
u(1-2\nu)}\right]  du \right)
$$
\begin{equation} \label{callprice2}
= P(0,T) \left(R_{\nu} +  \frac{1}{\pi} s_0^{\nu} K^{1-\nu}
\sum_{j=1}^M \mathbb{I}_{\alpha(0)=j} \int_{0}^{+ \infty}
\Re\left[ \frac{\e^{-\ci u \log(s_0/K) } \mathrm{q}_j^{0,T}(-u-\ci
\nu)}{\nu^2-\nu -u^2 + \ci u(1-2\nu)}\right] du \right),
\end{equation}
$$
R_\nu  =  \left \{
\begin{array}{ll}
                   0 & \nu > 1 \\
                   s_0 \frac{\varphi_T(-\ci)}{2} & \nu=1 \\
                   s_0 \varphi_T(-\ci) & 0 < \nu < 1 \\
                   s_0 \varphi_T(-\ci) - \frac{\e^k}{2} & \nu=0 \\
                   s_0 \varphi_T(-\ci) -\e^k &  \nu < 0\\
                 \end{array} \right.
$$
where $\varphi_T(-\ci)= \E[\e^{\sum_{i=1}^m \mu_i T_i}]$ according
to Remark (\ref{remark1}) and the functions $\mathrm{q}_i^{0,T}
(\cdot)$ are defined in (\ref{charfun3}).

\paragraph{Call/Put value in $\log$-strike transform.}
As before, if the GFT $\phi_j(\cdot)$ are well defined functions
in a properly defined strip of $\mathbb{C}$, from formula
(\ref{transfprices}) and Table \ref{GFTab}, we get for the call
option
$$
\E^{\mathcal{Q}}[\hat \Pi_{X(T)+\log(s_0)}(z)] =
\frac{\E[\e^{(X(T)+\log(s_0))(\ci z +1)}]}{\ci z-z^2} =
\frac{\e^{\log(s_0)(1+\ci z)} \varphi_T(z-\ci)}{\ci z-z^2} \ \ \
\Im(z) < 0
$$
from which
\begin{equation} \label{callprice_logstrike}
C_0 = \frac{P(0,T)}{2 \pi} \int_{\ci \nu-\infty}^{\ci \nu+\infty}
\e^{-\ci z k} \frac{\e^{\log(s_0)(1+\ci z) } \varphi_T(z-\ci)}{\ci
z-z^2} dz \ \ \ \nu < 0.
\end{equation}
$$
= \frac{P(0,T)}{2 \pi} \e^{(1-\nu)\log(s_0) + \nu k}
\int_{-\infty}^{+\infty} \e^{\ci u (k - \log(s_0))}
\frac{\varphi_T(u+\ci(\nu-1))}{\nu^2-\nu - u^2 + \ci u (1-2 \nu)}
du
$$
$$
= \frac{P(0,T)}{2 \pi} s_0^{1-\nu} K^\nu \int_{-\infty}^{+\infty}
\e^{- \ci u \log(K/s_0))} \frac{\varphi_T(u+\ci(\nu-1))}{\nu^2-\nu
- u^2 + \ci u (1-2 \nu)} du.
$$

The value for the put option and the related put-call parity are
obtained again by moving the integration contour. Since the
residues at the poles $z=0$ and $z=\ci$ of the integrand are
$\frac{s_0 \varphi_T(-\ci)}{\ci}$ and $\ci \e^k$ respectively, the
application of the residue Theorem gives the following general
formula for the call price in our RSJD model:
\begin{equation}
C_0(s_0,\alpha_0,K) = P(0,T)\left( R_\nu + \frac{1}{2 \pi}
\int_{\ci \nu-\infty}^{\ci \nu+\infty} \e^{-\ci z k}
\frac{\e^{\log(s_0)(1+\ci z) } \varphi_T(z-\ci)}{\ci z-z^2} dz
\right)
\end{equation}
$$
=P(0,T)\left( R_\nu + \frac{1}{\pi} s_0^{1-\nu} K^\nu
\int_{0}^{+\infty} \Re \left[ \e^{- \ci u \log(K/s_0))}
\frac{\varphi_T(u+\ci(\nu-1))}{\nu^2-\nu - u^2 + \ci u (1-2
\nu)}\right] du \right)
$$
\begin{equation} \label{callprice_logK}
= P(0,T)\left( R_\nu + \frac{1}{\pi} s_0^{1-\nu} K^\nu
\sum_{j=1}^M \mathbb{I}_{\alpha(0)=j} \int_{0}^{+\infty} \Re
\left[ \frac{\e^{- \ci u
\log(K/s_0)}\mathrm{q}_j^{0,T}(u+\ci(\nu-1))}{\nu^2-\nu - u^2 +
\ci u (1-2 \nu)}\right] du \right)
\end{equation}
$$
R_\nu = \left \{ \begin{array}{ll}
                  0  & \nu < 0 \\
                  s_0 \frac{\varphi_T(-\ci)}{2} & \nu = 0  \\
                  s_0 \varphi_T(-\ci)  & 0 < \nu < 1 \\
                  s_0 \varphi_T(-\ci) - \frac{\e^k}{2}  & \nu = 1 \\
                  s_0 \varphi_T(-\ci) - \e^k & \nu > 1
                 \end{array} \right.
$$

{\rem Let us notice that due to the symmetry of the call payoff,
the two approaches give in general very similar pricing formulas:
in particular from (\ref{callprice_logprice}) and
(\ref{callprice_logstrike}) it follows that by changing $z$ with
$\ci-z$ we can switch from one representation to the other.}

\begin{table}[t]
\begin{center}
\begin{tabular}{c|c}
  \hline
  Payoff & Option value in log-price transform\\ \hline \\
  $(\e^k-\e^y)^+$ & $P(0,T) \frac{1}{\pi} s_0^{\nu} \e^{k(1-\nu)}
\sum_{j=1}^M \mathbb{I}_{\alpha(0)=j} \int_{0}^{+ \infty}
\Re\left[ \frac{\e^{-\ci u (\log(s_0) - k)}
\mathrm{q}_j^{0,T}(-u-\ci
\nu)}{\nu^2-\nu -u^2 + \ci u(1-2\nu)} \right] du, \ \  \nu < 0$ \\
\\ \hline \\
  $\e^{a y} \mathbb{I}_{by>\kappa}$ & $P(0,T) \frac{1}{\pi} s_0^{\nu} \e^{(a-\nu)k/b}
  \sum_{j=1}^M \mathbb{I}_{\alpha(0)=j} \int_0^{+\infty} \Re \left [ \frac{\e^{-\ci u (\log(s_0)-k/b)}
  \mathrm{q}^{0,T}_j(-u-\ci \nu)}{\nu-a-\ci u}  \right] du, \ \ \nu > a$
  \\ \\
  \hline \\
  $\min (\e^y, \e^k)$ & $P(0,T)\frac{1}{\pi} s_0^{\nu} \e^{k(1-\nu)}
  \sum_{j=1}^M \mathbb{I}_{\alpha(0)=j} \int_0^{+\infty} \Re \left [ \frac{\e^{-\ci u
  (\log(s_0)-k)}
  \mathrm{q}^{0,T}_j(-u-\ci \nu)}{u^2-\nu^2 + \nu + \ci u (2 \nu-1)} \right] du, \ \ 0 < \nu < 1$ \\
  \hline
  & Option value in log-strike transform\\ \hline \\
  $(\e^k-\e^y)^+$ & $P(0,T) \frac{1}{\pi} s_0^{1-\nu} \e^{k \nu}
 \sum_{j=1}^M \mathbb{I}_{\alpha(0)=j} \int_{0}^{+\infty} \Re
 \left[ \frac{\e^{- \ci u (k-\log(s_0))}\mathrm{q}_j^{0,T}(u+\ci(\nu-1))}{\nu^2-\nu - u^2 +
 \ci u (1-2 \nu)}\right] du, \ \  \nu > 1$ \\ \\
\hline \\
  $\e^{a y} \mathbb{I}_{by>k}$ & $P(0,T) \frac{1}{\pi} s_0^{a-b\nu} \e^{\nu k}
  \sum_{j=1}^M \mathbb{I}_{\alpha(0)=j} \int_0^{+\infty} \Re \left [ \frac{\e^{-\ci u (k- b \log(s_0))}
  \mathrm{q}^{0,T}_j(a- \nu b + \ci u b)}{\ci u - \nu}  \right] du, \ \ \nu > 0$
  \\ \\
  \hline \\
  $\min (\e^y, \e^k)$ & $P(0,T)\frac{1}{\pi} s_0^{1-\nu} \e^{k \nu}
  \sum_{j=1}^M \mathbb{I}_{\alpha(0)=j} \int_0^{+\infty} \Re \left [ \frac{\e^{-\ci u
  (k-\log(s_0))} \mathrm{q}^{0,T}_j(u+\ci (\nu-1))}{u^2-\nu^2 + \nu + \ci u (2 \nu-1)}
  \right] du, \ \ 0 < \nu < 1$ \\ \\ \hline
\end{tabular}
\caption{Option values for some typical payoffs under the RSJD
model. See the Appendix for a sketch of their derivation.}
\label{TABVLOGP}
\end{center}
\end{table}

\paragraph{Application of the FFT algorithm.} As it is widely
known, the transform method deserves for an efficient evaluation
of derivative prices by means of the FFT algorithm for a proper
range of the trigger parameter. Actually, if only one option price
has to be evaluated for a fixed $k$, there is no need to use FFT.
This technique involves two steps:
\begin{enumerate}
\item a numerical quadrature scheme to approximate the integral
appearing in the pricing formula, that we write as
$$
I(k)=\frac{1}{\pi} \int_{0}^{+\infty}  \Re \left[ \e^{-\ci u k}
F(u)\right] du,
$$
through a $N$-point sum. By using an equispaced grid $\{ u_n
\}_{n=1,\ldots,N}$  of the line $\{z=u + \ci v \in \mathbb{C}:
u\in \mathbb{R}^+, v=\nu \}$ with spacing $\Delta$, we have
$$
I(k) \approx \Sigma_N(k) = \frac{\Delta}{\pi}\sum_{n=1}^N \Re\left
[ \e^{-\ci u_n k} F(u_n) w_n \right],
$$
where $w_n$ are the integration weights;
\item given a properly spaced grid of triggers $k_m = k_1 + \gamma
(m-1)$, $m=1,\ldots N$, the sum $\Sigma_N(k)$ is written as a
discrete Fourier transform (DFT), so that the FFT algorithm can be
used.
\end{enumerate}

\paragraph{A numerical example.}

In order to asses the performances of the pricing formulas we
consider the basic models in Remark (\ref{remark2}), Example
(\ref{ex_2rsjdmerton}), in which the regime switching behavior is
driven by a two-state Markov chain. We fit these models on a set
of observed call prices on the S\&P 500 index as quoted on March
31, 2009 to get realistic values for the parameters. In the data
set used for calibrating the models there are $128$ call option
prices with maturities and strike prices ranging from $31$ to
$272$ days and from $525$ to $1200$, respectively. The value of
the index is $s_0=753.89$ and the moneyness $s_0/K$ ranges from
$0.6282$ to $1.4360$. The average of the bid and ask Treasury bill
discounts, as available from the \textit{Wall Street Journal},
were used and converted to annualized risk-free rates. The
dividend rate $q$ was estimated from the data: in particular we
used a non linear least squares algorithm which minimize the
difference between observed call prices and the corresponding
Black \& Scholes prices evaluated through the available implied
volatility, constrained to satisfy the put-call parity relations.
This procedure was repeated for each maturity giving a mean value
$q=0.0157$ with standard deviation $0.003$.

\medskip

\begin{figure}[t]
\begin{center}
\includegraphics[width=12cm,height=8cm]{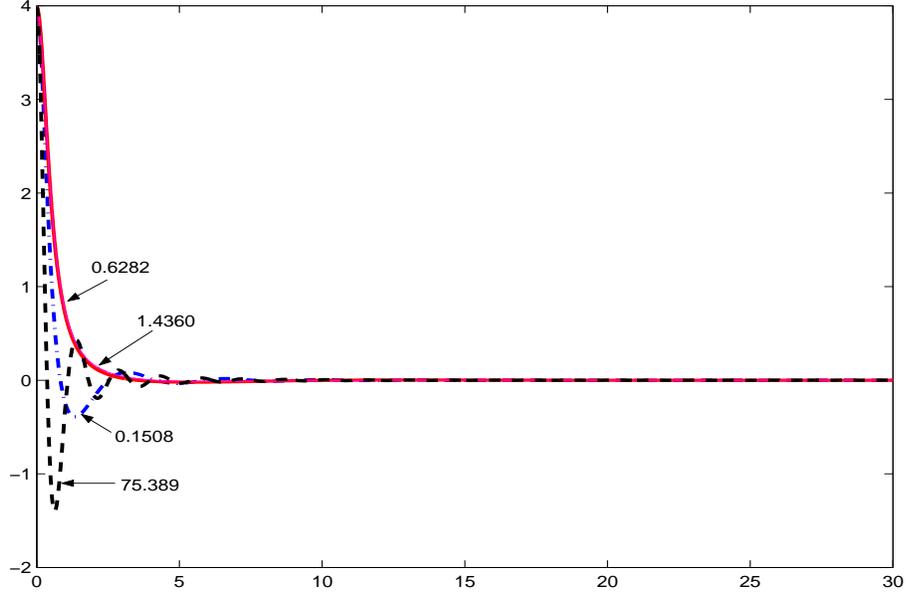}
\caption{\small Real part of the integrand for a call option for
different values of the moneyness. The plots related to our data
($0.6282$ and $1.4360$) are very similar.} \label{integ_fig}
\end{center}
\end{figure}

The numerical implementation was developed in the
MatLab$^\copyright$ environment. Quadrature algorithms are needed
to evaluate the option prices from (\ref{callprice2}): adaptive
Simpson and Gauss-Lobatto quadrature rules, as available in
MatLab, performed equally well, for typical values of the
parameters. As a matter of fact the integrands are not rapidly
oscillating and decrease sufficiently fast (e.g. see Fig.
(\ref{integ_fig})). The FFT algorithm was implemented following
Lee (2004), i.e. by sampling $F$ at the midpoints of intervals of
length $\Delta$, $u_n = (n-\frac{1}{2}) \Delta$, $n=1,\ldots, N$
and taking $\gamma \Delta = 2 \pi / N $. We get
$$
\Sigma_N(k_m) = \frac{\Delta}{\pi} \Re\left[ \sum_{n=1}^N \e^{-\ci
(n-\frac{1}{2}) \Delta (k_1 + \gamma(m-1))} F((n-\frac{1}{2})
\Delta) \right]
$$
$$
=   \frac{\Delta}{\pi} \Re\left[ \e^{-\ci \frac{\pi}{N} (m-1)}
\sum_{n=1}^N \e^{-\ci \frac{2 \pi}{N} (n-1)(m-1)} f(n) \right]
$$
where $f(n) = F((n-\frac{1}{2}) \Delta) \e^{-\ci (n-\frac{1}{2})
\Delta k_1}$. In this case we used $\gamma = 0.01$ and $\nu=0.5$.


\medskip

For the calibration we minimized the sum of squared errors by
using the constrained minimization routine in MatLab. In fact, for
the regime switching models we have to add the constraint
$\sigma_1 > \sigma_2$. The results obtained are reported in Table
(\ref{TAB1}): RMSE and relative errors $\frac{\hat C -
C_0(s_0,K)}{\hat C}$ were calculated in the four cases. In Table
(\ref{TAB2}) we report out-of-sample performances of each fitted
model: these were obtained by calculating the deviation from five
call option prices having a much longer maturity, i.e. $631$ days
and moneyness ranging from $0.7539$ to $1.0052$.

\begin{figure}[t]
\begin{center}
\includegraphics[width=12cm,height=10cm]{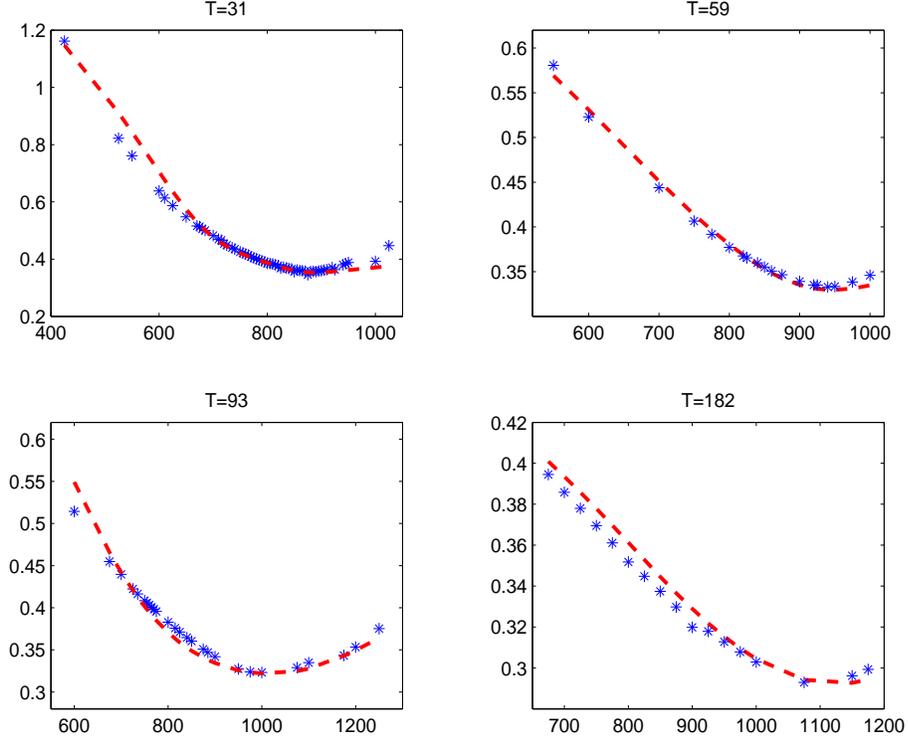}
\caption{\small Implied volatility calibration. } \label{fig025}
\end{center}
\end{figure}

\begin{table}[h]  
\begin{tabular}{c|c|c|c|c}
              & BS     & RSBS   & JDM & RSJDM \\
  \hline
  $\sigma_1$  & 0.3645 & 0.4462 & 0.1341  &  0.2725  \\
  $\sigma_2$  &        & 0.3296 &         &  0.1350  \\
  $\lambda_1$ &        &        & 7.9958  &  6.8393  \\
  $\lambda_2$ &        &        &         &  0.8590  \\
  $a_1$       &        &        & -0.1280 &  -0.1398 \\
  $a_2$       &        &        &         &  -0.3423  \\
  $b_1$       &        &        & 0.0011  &  0.0877  \\
  $b_2$       &        &        &         &  0.1593  \\
  $q_1$       &        & 9.6199 &         &  6.5075  \\
  $q_2$       &        & 0.0002 &         &  0.0020  \\   \hline
  RMSE        & 4.6947 & 3.9177 ($i_0=1$) & 3.8715  & 0.6126 ($i_0=1$) \\
  Rel. err. & \begin{tabular}{c}
    (-0.0242,0.1398) \\
    0.0888\\
  \end{tabular}   & \begin{tabular}{c}
    (-0.0141, 0.0631) \\
    0.0353 \\
    \end{tabular} & \begin{tabular}{c}
      (-0.0508, 0.1547) \\
      0.0675 \\
    \end{tabular}
  & \begin{tabular}{c}
    (-0.0080, 0.0191) \\
    0.0041 \\
  \end{tabular}
\end{tabular}
\caption{Implied parameters and in-sample calibration performances
of the models. The moneyness of these options ranges from $0.6282$
to $1.4360$. In the last rows we report the range and the mean of
the relative pricing error $\frac{\hat C - C_M}{\hat
C}$.}\label{TAB1}
\end{table}

\bigskip

\begin{table}[h]  \centering
\begin{tabular}{c|c|c|c|c}
              & BS     & RSBS   & JDM & RSJDM \\ \hline
RMSE          & 14.8555  &   6.4732  & 17.1020  & 5.1116 \\
Mean Rel Err. &  -0.1852 & -0.0631 & -0.2142 & -0.0634

\end{tabular}
\caption{Out-of-sample performance of the models. The moneyness of
these options ranges from $0.7539$ to $1.0052$.}\label{TAB2}
\end{table}

\section{On the pricing of forward starting options}

Forward starting options are well-known exotic derivatives,
depending on an underlying asset characterized by the payoff
\begin{equation} \label{fso_payoff}
\Pi_T(S(T),\kappa) = S(T) - \kappa S(t^*),
\end{equation}
where $t^* \in (0, T)$ is the determination time and $\kappa \in
(0,1)$ is a given percentage. They are the building blocks of the
so-called cliquet options and are used in many different context.

In this Section we provide a simple valuation formula for the
price at time $t=0$ of this claim where the underlying $S(t)$
follows the regime-switching jump diffusion dynamic introduced in
Sect. 2. Furthermore, we assume that $\mu(\alpha)=r$, the
risk-free rate, in such a way $\mathcal{P}\equiv\mathcal{Q}$. The
risk-neutral price is therefore given by
$$
\Pi_0(s_0,\alpha_0,\kappa) = \E[\e^{-r T}(S(T)-\kappa S(t^*))^+] =
\E[\e^{-r T}(s_0\e^{X(T)}-\kappa S(t^*))^+].
$$

Notice that in general from the determination time $t^*$ on, the
price is equal to that of a standard call option, being the strike
a known constant. By denoting with $\E_t[\cdot]$ the conditional
expectation w.r.t. information up to time $t$, $\mathcal{F}_t$, we
have
$$
C_t(S(t),\alpha(t),K) = \E_t[\e^{-r (T-t)}(S(T)-K)^+] =
\e^{-r(T-t)} \E_t[(S(t)\e^{X_{t,T}}-K)^+].
$$

Therefore, if $K =\kappa S(t)$ we get that
$$
C_t(S(t),\alpha(t),\kappa S(t)) = S(t) C_t(1,\alpha(t), \kappa).
$$

Hence, by the law of iterated conditional expectations,
$$
\Pi_0(s_0,\alpha_0,\kappa) = \E[\e^{-r
T}(S(t^*)\e^{X_{t^*,T}}-\kappa S(t^*))^+] =
\E[\e^{-rt^*}C_{t^*}(S({t^*}),\alpha({t^*}),\kappa S({t^*}))] =
$$
$$
=  \E[S(t^*)\e^{-rt^*} C_{t^*}(1,\alpha(t^*), \kappa)].
$$

Since $S(t)\e^{-rt}=S(t)/B(t)$ is a $\mathcal{Q}$-martingale we
can introduce an equivalent measure $Q^S$ as
$$
L(t)=\frac{dQ^S}{dQ} |_t = \frac{S(t)}{B(t)} \frac{B(0)}{s_0}, \ \
\E[L(T)]=1,
$$
from which we get
$$
\E[S(t^*)\e^{-rt^*} C_{t^*}(1,\alpha(t^*), \kappa)] = s_0
\E^{Q^S}[C_{t^*}(1,\alpha(t^*), \kappa)].
$$

Notice that this property is fairly general: in fact $\alpha(t)$
is not restricted to be a Markov chain. On the other hand, in our
model we don't need to further specify the $Q^S$ dynamic of the
price process, since the Markov chain is not affected by this
change of measure. Hence, since the chain is assumed to be
stationary, by denoting with $\mathbf{\pi} = (\pi_1, \ldots,
\pi_M)'$ its invariant probability, we get
$$
\Pi_0(s_0,\kappa) = s_0 \E^{Q^S}[C_{t^*}(1,\alpha(t^*), \kappa)] =
s_0 \sum_{j=1}^M \pi_j C_{t^*}(1,j,\kappa).
$$

The price of the forward starting option is therefore the mixture
of call option prices evaluated at the determination time $t^*$
under each regime weighted by the corresponding probability. By
using transform representation for the call option value, e.g.
(\ref{callprice2}), we get the following proposition.

{\proposition \label{FSOPrice} Let the underlying be characterized
by the SDE (\ref{rsjdm_sde}). Then the price at time $t=0$ of the
Forward Starting Option (\ref{fso_payoff}) with determination time
$t^*$ and percentage $\kappa \in (0,1)$ is given by
$$
\Pi_0(s_0,\kappa) = s_0 \e^{-r(T- t^*)} \sum_{j=1}^M \pi_j
\left(R_{\nu} +  \frac{ \kappa^{1-\nu}}{\pi} \int_{0}^{+ \infty}
\Re\left[ \frac{\e^{-\ci u \log(\kappa)}
\mathrm{q}_j^{t^*,T}(u-\ci \nu)}{\nu^2-\nu -u^2 - \ci
u(1-2\nu)}\right] du \right),
$$
where $\{ \pi_j \}_{j=1,\ldots,m}$ is the stationary probability
of the Markov chain, $\nu$ and $R_\nu$ following from
(\ref{callprice2}). $\Box$ }

\bigskip

As a byproduct of the last proposition, by restricting our model
to a unique regime (see Remark (\ref{remark2})), we get a simple
formula for pricing a FSO in a L\'{e}vy model with finite
activity.

The impact of model choice on the prices of the Forward Starting
options is shown in figures (\ref{fig025}), (\ref{fig05}) and
(\ref{fig075}) as a function of the determination time for three
different values of the percentage $\kappa$. The parameters of
each model are those estimated in our numerical example (Table
(\ref{TAB1})).

\begin{figure}[h]
\begin{center}
\includegraphics[width=10cm,height=8cm]{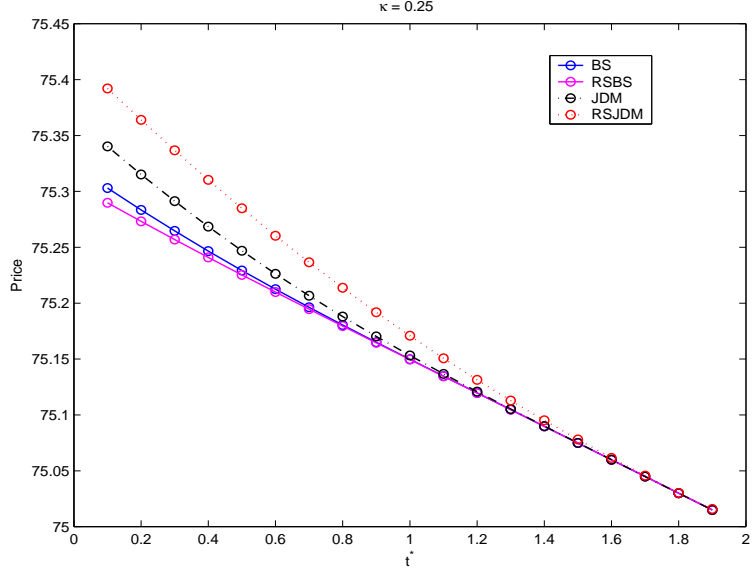}
\caption{\small Forward starting option prices. } \label{fig025}
\end{center}
\end{figure}

\begin{figure}[h]
\begin{center}
\includegraphics[width=10cm,height=8cm]{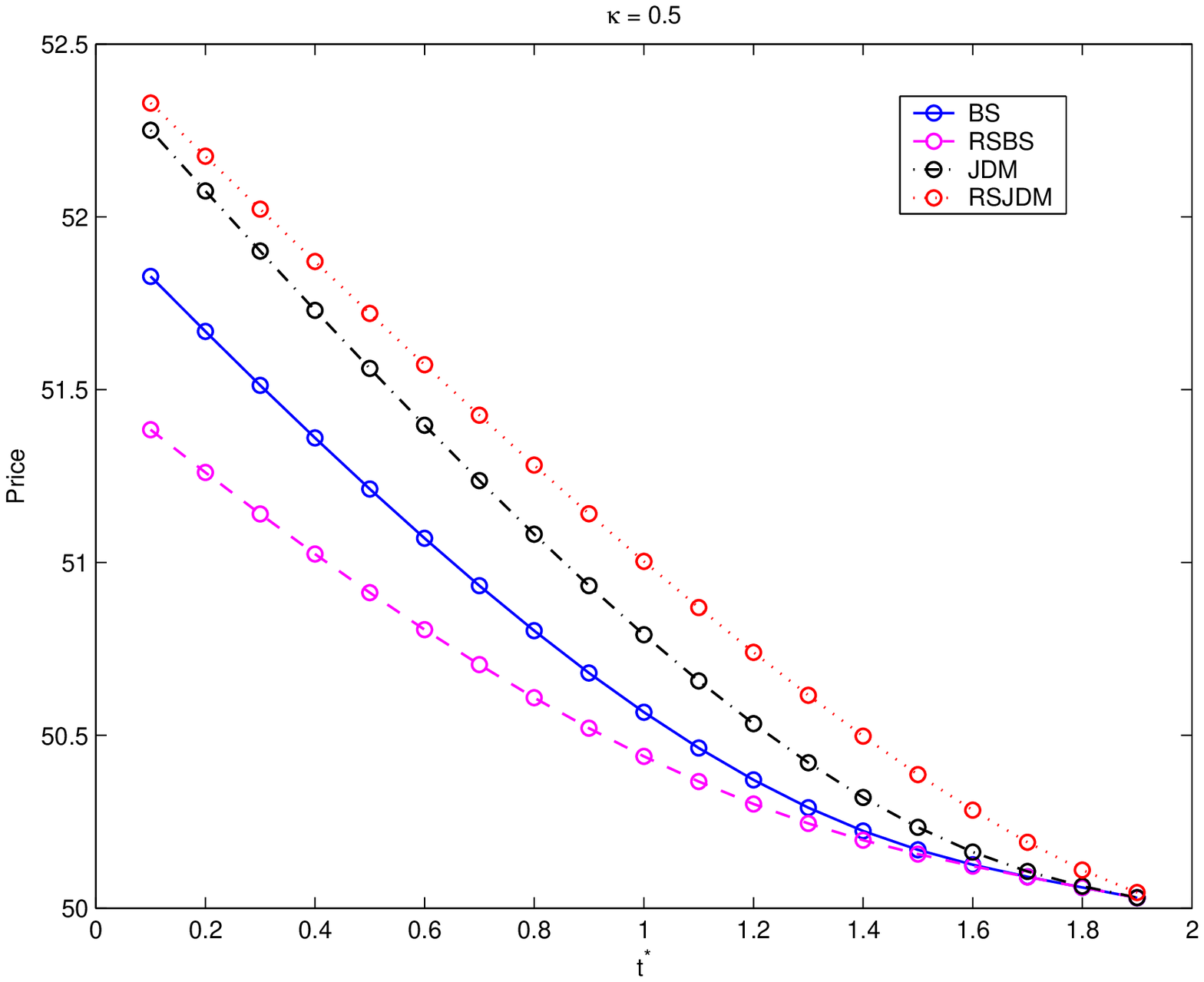}
\caption{\small Forward starting option prices. } \label{fig05}
\end{center}
\end{figure}

\begin{figure}[h]
\begin{center}
\includegraphics[width=10cm,height=8cm]{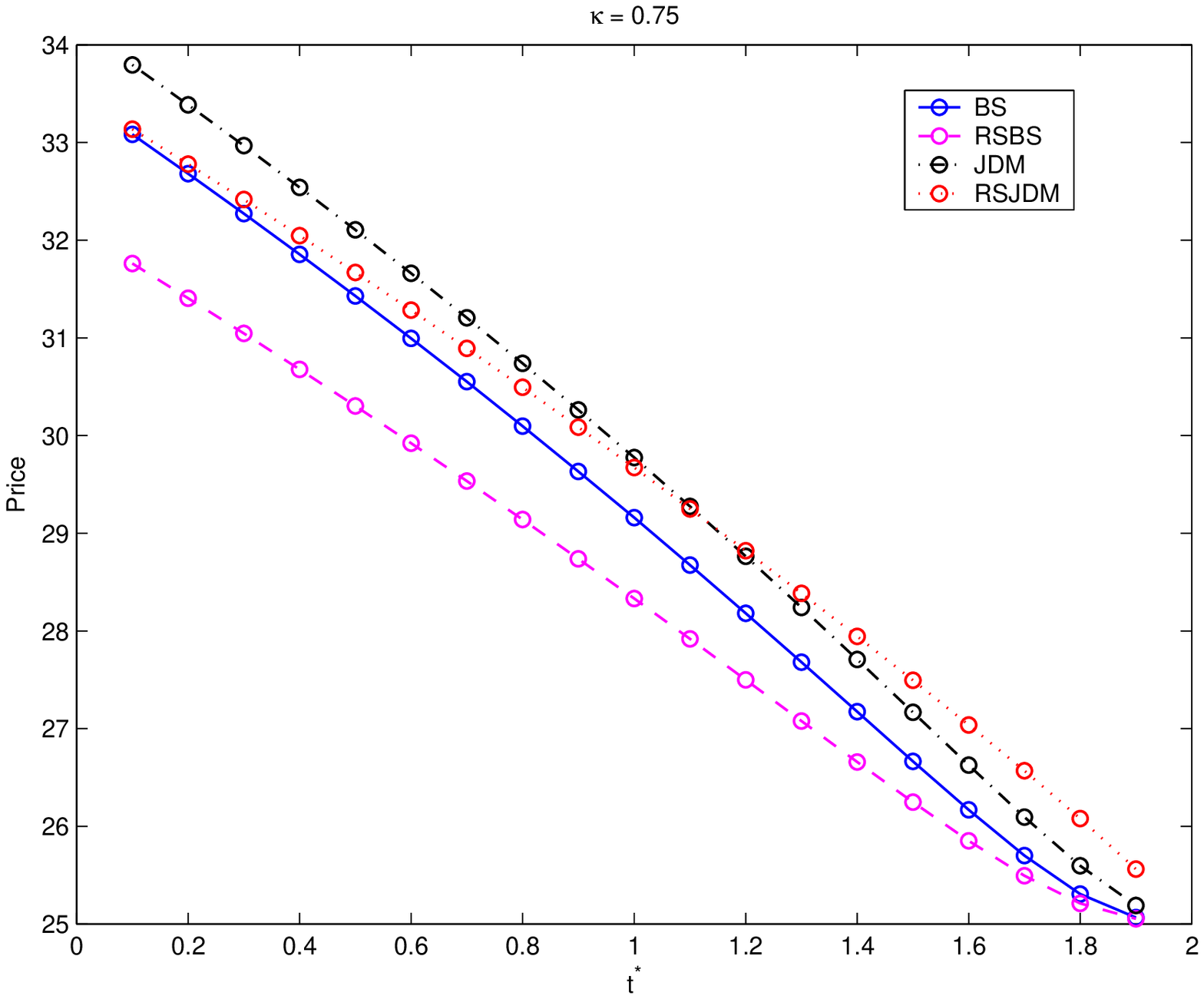}
\caption{\small Forward starting option prices. } \label{fig075}
\end{center}
\end{figure}

\bigskip

\paragraph{Pricing FSO in a general stochastic volatility model.}
In Chourdakis (2004) a regime-switching diffusion was considered
to approximate a general stochastic volatility model
\begin{eqnarray}
dX(t) & = &  \mu(v(t)) dt + \sigma(v(t)) dW(t) + \int_\mathbb{R} y p^{\alpha}(dy,dt), \\
dv(t) & = & a(v(t)) dt + b(v(t)) dZ(t) \label{volproc},
\end{eqnarray}
where $X(t) = \log(S(t))$ and $p^{\alpha}(dy,dt)$ has intensity
$\lambda(t,v,dy)= \lambda(v) m_{v}(y) dy$, $m_v$ being the
probability measure which characterizes the jump component. Then,
under some conditions on the coefficients $a(\cdot)$ and
$b(\cdot)$, the diffusion process (\ref{volproc}) can be
approximated by a finite state Markov chain defined on a grid
$G^{\epsilon}$ which is the discretization of the domain of $v$.
The approximating scheme defines a generator $Q^{\epsilon}$ for
the Markov chain depending on $\epsilon$ and on the functions
$a(\cdot)$ and $b(\cdot)$ evaluated at the points of the grid
$G^{\epsilon}=\{v^\epsilon_1, \ldots, v^\epsilon_M\}$. As reported
in Chourdakis (2004) the generator $Q^{\epsilon} = \{
q^\epsilon_{ij} \}_{i,j=1,\ldots,M}$ where
\begin{equation} \label{Qapprox}
q^\epsilon_{ij} = \left\{ \begin{array}{ll}
 \frac{1}{2\epsilon^2} b^2(v^\epsilon_j) - \frac{1}{2\epsilon} a(v^\epsilon_j),& i=j-1
 \\ \\
 \frac{1}{\epsilon^2} b^2(v^\epsilon_j),                                      & i=j
 \\ \\
 \frac{1}{2\epsilon^2} b^2(v^\epsilon_j) + \frac{1}{2\epsilon} a(v^\epsilon_j), & i=j+1
 \\ \\
                            0 & i \neq j-1,j,j+1 \\
                          \end{array} \right.
\end{equation}
produces accurate results for coarse volatility grids. The
resulting approximated process $X^{\epsilon}(t)$ follows therefore
a RSJD dynamic and its characteristic function is obtained from
Prop. (\ref{propCFX}). Correspondingly, option prices can be
calculated by means of the Fourier transform techniques presented
in Sect. 3. Convergence properties as well as computational
considerations as $\epsilon \rightarrow 0$ are discussed in
Chourdakis (2004) where a number of cases are studied.

This technique combined with our Proposition (\ref{FSOPrice})
suggests the following scheme for pricing FSO under a general SV
model:
\begin{enumerate}
\item approximate the model with a regime-switching diffusion $X^{\epsilon}(t)$:
this amounts to build the generator $Q^{\epsilon}$ of the Markov
chain, defined by (\ref{Qapprox});
\item evaluate call option prices  $C^{\epsilon}_j$ at time $t^*$ under each regime
$j=1,\ldots,M$ through formula (\ref{callprice2}) or
(\ref{callprice_logK}) with $S(t^*)=1$ and $K=\kappa$;
\item calculate the price $\Pi_0(s_0,\kappa)= s_0 \sum_{j=1}^M \pi_j C^{\epsilon}_j$
where the coefficients $\pi_j$'s are the solution of $\pi
Q^{\epsilon} = 0$.
\end{enumerate}

\section{Conclusion}

In this paper we considered the problem of valuing the price of a
European contingent claim when the underlying dynamic follows a
L\`{e}vy process of I type whose parameters are modulated by a
continuous time and finite state Markov chain. These kind of
processes are known to capture specific features of financial time
series, such as volatility clustering and structural breaks. On
the other hand, they can equally be used to approximate very
general stochastic volatility processes. Following the well
established relationship between option prices and Fourier
transforms, we obtained almost closed-form solutions (up to a
numerical integration) for European style options, both in
log-price and in log-strike space. An example of calibration for
the regime-switching version of the Merton jump-diffusion model is
also presented for a daily set of call option data on the S\&P
500.

Furthermore, as a practical application of the Fourier transform
methodology we obtained an almost closed-form solution to the
problem of valuing a Forward Starting option in our general
regime-switching jump-diffusion dynamic. This result can be
jointly used with the approximation scheme of stochastic
volatility models to get a feasible algorithm for FSO pricing
under a very general dynamic.

\section{Appendix}

\paragraph{Proof of \ref{Xrepres}}
Let us define the occupation times for the Markov chain
$\alpha(t)$ in $[0,T]$, $T_i = \int_0^T \mathbb{I}_{\alpha(s)=i}
ds$, $i=1,\ldots,M$. We immediately have that $\sum_{i=1}^m
T_i=T$. Now, given a sample path of the chain $\alpha(t), 0 \leq t
\leq T$, we can define
\begin{eqnarray}
\label{delta_i} \Delta_i & = & \bigcup_{\ell: \  \alpha(t)=i,
\tau_\ell \leq t < \tau_{\ell+1}} [\tau_\ell , \tau_{\ell+1}) \\
\label{poiss_i} N(\Delta_i) & = & \sum_{\ell: \ \alpha(t)=i,
\tau_\ell \leq t < \tau_{\ell+1}} \left( N_{\tau_{\ell+1}} -
N_{\tau_{\ell}}
\right ) \\
\label{norm_i} Z(\Delta_i) & = & \sum_{\ell: \ \alpha(t)=i,
\tau_\ell \leq t < \tau_{\ell+1}} \left( W(\tau_{\ell+1}) -
W(\tau_{\ell}) \right).
\end{eqnarray}

Since each $\Delta_i$ is the union of non overlapping intervals,
the corresponding random variables $N(\Delta_i)$ and $Z(\Delta_i)$
are distributed as a Poisson variable $\mathrm{Poiss}(\lambda_i
T_i)$ and as a Normal variable $\mathcal{N}(0,T_i)$, respectively.
Furthermore, $N(\Delta_i) \perp N(\Delta_j)$ and $Z(\Delta_i)
\perp Z(\Delta_i)$, for $i \neq j$\footnote{Here, $X \perp Y$
means that $X$ and $Y$ are independent.}. By denoting with
$Y^{(i)}_k$ the $k$-th jump magnitude relative to regime $i$, we
have
\begin{equation} \label{jumps}
J(t) = \sum_{i=1}^M \sum_{k=1}^{N(\Delta_i)} Y^{(i)}_k \ \ \
\mbox{and} \ \ \int_0^T \sigma(\alpha(s)) dW(s) = \sum_{i=1}^m
\sigma(i) Z(\Delta_i) \sim \mathcal{N}(0, \sum_{i=1}^m \sigma_i^2
T_i)
\end{equation}

By defining $\xi(\alpha) =\mu(\alpha) - \frac{1}{2}
\sigma^2(\alpha) - \lambda(\alpha) \kappa(\alpha)$ and
\begin{eqnarray*}
\Xi_T(T_1, \ldots,T_m) = \int_0^T \xi(\alpha(s)) ds = \sum_{i=1}^m
\xi(i) T_i
\end{eqnarray*}
then $X(T)$ admits the following representation:
\begin{equation} \label{rsjd2}
X(T) = \Xi_T(T_1, \ldots,T_m) + \sum_{i=1}^m \sigma(i) Z(\Delta_i)
+\sum_{i=1}^M \sum_{k=1}^{N(\Delta_i)} Y^{(i)}_k.
\end{equation}

\paragraph{Proof of \ref{propCFX}}
Let $\phi_j(z)=\E[\e^{\ci z \gamma(Y(j),j)}]$ be the generalized
Fourier transform of the jump magnitude. From the representation
(\ref{rsjd2}) we can easily calculate the characteristic function
of $X(T)$, conditional to $\mathcal{F}^{\alpha}_T$:
$$
\E[\e^{\ci z X(T)} | \mathcal{F}^{\alpha}_T] = \e^{\ci z
\Xi_T(T_1, \ldots,T_m) } \ \E[\e^{\ci z \sum_{j=1}^m \sigma(j)
Z(\Delta_j)} | \mathcal{F}^{\alpha}_T] \ \E[\e^{\ci z \sum_{j=1}^M
\sum_{k=1}^{N(\Delta_j)} \gamma(Y^{(j)}_k,j)}|
\mathcal{F}^{\alpha}_T].
$$
The first expected value is simply obtained as
$$
\E[\e^{\ci z \sum_{j=1}^m \sigma(i) Z(\Delta_j)} |
\mathcal{F}^{\alpha}_T] = \e^{-\frac{1}{2} z^2 \sum_{j=1}^m
\sigma_j^2 T_j},
$$
while the second, since $N(\Delta_i) \perp N(\Delta_j)$ for $i
\neq j$, is
$$
\E[\e^{\ci z \sum_{j=1}^M \sum_{k=1}^{N(\Delta_j)}
\gamma(Y^{(j)}_k,j)}| \mathcal{F}^{\alpha}_T] = \e^{\sum_{j=1}^m
\lambda_j T_j (\phi_j(z)-1)}.
$$

Finally, we have
\begin{equation} \label{charfun1}
\varphi_T(z) = \E \left[ \e^{\ci z \Xi_T(T_1, \ldots,T_m) -
\frac{1}{2} z^2 \sum_{j=1}^m \sigma_j^2 T_j + \sum_{j=1}^m
\lambda_j T_j (\phi_j(z)-1) } \right].
\end{equation}

Actually, the exponent in (\ref{charfun1}) is a linear function of
the sojourn times $T_1, \ldots, T_m$, the characteristic function
of which are well-known. As a matter of fact, we have
\begin{equation} \label{charfun4}
\varphi_T(z) = \E \left[ \e^{\ci \sum_{j=1}^m \vartheta_j(z) T_j}
\right] = \e^{\ci \vartheta_m(z) T} \E \left[ \e^{\ci
\sum_{j=1}^{m-1} \tilde \vartheta_j(z) T_j} \right]
\end{equation}
where $\tilde \vartheta_i(z) = \vartheta_i(z) - \vartheta_m(z)$,
being $T_m = T -(T_1+\ldots+T_{m-1})$. Since it can be proved (see
e.g. Buffington and Elliott (2002)), that
$$
\E \left[ \e^{\ci \sum_{j=1}^{m-1} \tilde \vartheta_j T_j} \right]
= \mathbf{1}' \cdot \e^{Q'+\ci \mathrm{diag}(\tilde \vartheta_1,
\ldots, \tilde \vartheta_{m-1},0) T} \cdot \mathbb{I}(0),
$$
formula (\ref{charfun2}) follows, the second equality being a
consequence of the property of matrix exponential $\exp(\theta)
\exp(A) = \exp(\theta I +A)$.

\paragraph{Derivation of Tables \ref{TABVLOGP}.} Payoff
$\e^{a x} \mathbb{I}_{bx>\kappa}$. From the third row of Table
\ref{GFTab} and formula (\ref{transfprices}) we get for log-price
transform
$$
\Pi_0 / P(0,T) = \frac{1}{2 \pi} \int_{\ci \nu - \infty}^{\ci \nu
+ \infty} \e^{-\ci z \log(s_0)} \frac{\e^{(a+\ci z) k /b}}{a + \ci
z} \varphi_T(-z) dz =
$$
$$
\frac{1}{2 \pi} s_0^{\nu} \e^{(a - \nu) k /b}
\int_{-\infty}^{+\infty} \frac{\e^{-\ci u (\log(s_0)-k/b)}
\varphi_T(-u-\ci \nu)}{\nu-a-\ci u} du =
$$
$$
\frac{1}{\pi} s_0^{\nu} \e^{(a - \nu) k /b} \int_{0}^{+\infty} \Re
\left [\frac{\e^{-\ci u (\log(s_0)-k/b)} \varphi_T(-u-\ci
\nu)}{\nu-a-\ci u} \right ] du, \ \ \ \nu > a;
$$

for log-strike transform
$$
\E[\hat{\Pi}_{X(T)+\log(s_0)}(k)] = \E[\frac{\e^{(a+\ci z
b)(\log(s_0)+X(T))}}{\ci z}] = \frac{\e^{(a+\ci z
b)\log(s_0)}}{\ci z} \varphi_T(a+\ci z b)
$$
from which
$$
\Pi_0 / P(0,T) = \frac{1}{2 \pi} \int_{\ci \nu - \infty}^{\ci \nu
+ \infty} \e^{-\ci z k} \frac{\e^{(a+\ci z k) \log(s_0)}}{\ci z}
\varphi_T(a+\ci b z ) dz =
$$
$$
\frac{1}{2 \pi} s_0^{a- b \nu} \e^{\nu k} \int_{- \infty}^{+
\infty} \frac{\e^{-\ci u (k - b \log(s_0))}}{\ci u - \nu}
\varphi_T(a - \nu b + \ci b u ) du =
$$
$$
\frac{1}{\pi} s_0^{a- b \nu} \e^{\nu k} \int_{0}^{+ \infty} \Re
\left [\frac{\e^{-\ci u (k - b \log(s_0))}}{\ci u - \nu}
\varphi_T(a - \nu b + \ci b u ) \right ] du, \ \ \ \nu > 0.
$$

Payoff $\min (\e^x, \e^k)$. From the fourth row of Table
\ref{GFTab} and formula (\ref{transfprices}) we get for log-price
transform
$$
\Pi_0 / P(0,T) = \frac{1}{2 \pi} \int_{\ci \nu - \infty}^{\ci \nu
+ \infty} \e^{-\ci z \log(s_0)} \frac{\e^{k(\ci z +1)}}{z^2-\ci z}
\varphi_T(-z) dz =
$$
$$
\frac{1}{2 \pi} s_0^{\nu} \e^{k(1-\nu)} \int_{- \infty}^{+ \infty}
 \frac{\e^{-\ci u (\log(s_0)-k)}}{u^2-\nu^2+\nu+\ci u (2\nu-1)}
\varphi_T(-u-\ci \nu) du =
$$
$$
\frac{1}{\pi} s_0^{\nu} \e^{k(1-\nu)} \int_{- \infty}^{+ \infty}
\Re \left [ \frac{\e^{-\ci u (\log(s_0)-k)}}{u^2-\nu^2+\nu+\ci u
(2\nu-1)} \varphi_T(-u-\ci \nu) \right ] du, \ \ \ 0 < \nu < 1;
$$
for log-strike transform
$$
\E[\hat{\Pi}_{X(T)+\log(s_0)}(k)] = \E[\frac{\e^{(1+\ci
z)(\log(s_0)+X(T))}}{z^2-\ci z}] = \frac{\e^{(1+\ci z)
\log(s_0)}}{z^2-\ci z} \varphi_T(z-\ci)
$$
from which
$$
\Pi_0 / P(0,T) = \frac{1}{2 \pi} \int_{\ci \nu - \infty}^{\ci \nu
+ \infty} \e^{-\ci z k} \e^{-\ci z k} \frac{\e^{(1+\ci z)
\log(s_0)}}{z^2-\ci z} \varphi_T(z-\ci) dz =
$$
$$
\frac{1}{2 \pi} \e^{\nu k} s_0^{1-\nu} \int_{- \infty}^{+ \infty}
\frac{\e^{-\ci u (k-\log(s_0))}}{u^2-\nu^2+\nu + \ci u (2 \nu-1)}
\varphi_T(u+\ci (\nu-1)) du =
$$
$$
\frac{1}{\pi} \e^{\nu k} s_0^{1-\nu} \int_{0}^{+ \infty} \Re
\left[\frac{\e^{-\ci u (k-\log(s_0))}}{u^2-\nu^2+\nu + \ci u (2
\nu-1)} \varphi_T(u+\ci (\nu-1)) \right ] du, \ \ \ 0 < \nu < 1.
$$

By substituting in the previous formulas the GFT (\ref{charfun3})
of the RSJD model, we immediately get the entries of Table
\ref{TABVLOGP}.

\section*{Acknowledgments} The financial support of the Research Grant:
\emph{PRIN 2008, Probability and Finance, Prot. 2008YYYBE4}, is
gratefully acknowledged. Moreover, the author would like to thank
the participants to the workshop "Stochastic Volatility, Affine
Models and Transform Methods" organized by Prof. S. Herzel at the
School of Economics of the University of Roma - Tor Vergata, 15-16
April 2010.


\section*{References}

\begin{description}
\item[] Bakshi G. and Madan D. (2000), Spanning and
derivative-securities valuation, \emph{Journal of Financial
Economics}, 55, pp. 205-238.

\item[] F. Biagini, Y. Bregman, and T.Meyer-Brandis, Pricing of
catastrophe insurance options written on a loss index with
reestimation, \emph{Insurance Math. Econom.}, 43 (2008), pp.
214-222.

\item[] Bollen N. P. B. (1998), Valuing options in
regime-switching models. \emph{Journal of Derivatives}, 6, pp.
38-49.

\item[]  K. Borovkov and A. Novikov, On a new approach to calculating
expectations for option pricing, \emph{J. Appl. Probab.}, 39
(2002), pp. 889-895.

\item[] Boyarchenko S., Levendorskii S., American options in
regime-switching models. \emph{SIAM J. Control Optim.} 48 (2009),
no. 3, pp. 1353-1376.

\item[] Boyle P. and Draviam T. (2007), Pricing exotic
options under regime switching, \emph{Insurance: Mathematics and
Economics}, 40 , pp. 267-282.

\item[] Buffington J. and Elliott R. J. (2002), American options with
regime switching, \emph{International Journal of Theoretical and
Applied Finance}, 5 , pp. 497-514.

\item[] Carr P. and Madan D.B. (1999), Option valuation
using the Fast Fourier Transform, \emph{Journal of Computational
Finance}, 2, pp. 61-73.

\item[] Cherubini U., Della Lunga G., Mulinacci S.,
Rossi P., \textsl{Fourier Transform Methods in Finance}, Wiley,
2009.

\item[] Chourdakis K. (2004), Non-Affine option pricing,
\emph{The Journal of derivatives}, 2004, pp. 10-25.

\item[] Chourdakis K. (2007), L\'{e}vy process driven by stochastic volatility,
\emph{Asia-Pacific Financial Markets}, Vol. 12, No. 4, pp.
333-352.

\item[] Cont R. and Tankov P. (2004), \textsl{Financial Modelling with Jump
Processes}, Chapman \& Hall/CRC Financial Mathematics Series.

\item[] Di Graziano G. and Rogers L. C. G. (2009), Equity
with Markov-modulated dividends, \emph{Quantitative Finance}, Vol.
9, No. 1, pp. 19-26.

\item[] Di Masi, G.B., Kabanov, Y.M., Runggaldier,W.J. (1994),
Mean-variance hedging of options on stocks with Markov volatility,
\emph{Theory of Probability and Its Applications}, 39, pp.
173-181.

\item[] Duan, J.C., Popova, I., Ritchken, P. (2002), Option pricing under
regime switching, \emph{Quantitative Finance}, 2, pp. 1-17.

\item[] D. Dufresne, J. Garrido, and M. Morales, Fourier inversion
formulas in option pricing and insurance,\emph{ Methodol. Comput.
Appl. Probab.}, 11 (2009), pp. 359-383.

\item[] E. Eberlein, K. Glau, and A. Papapantoleon, Analysis of
Fourier transform valuation formulas and applications, \emph{Appl.
Math. Finance}, 17 (2010), pp. 211-240.

\item[] Edwards C. (2005), Derivative Pricing Models with Regime
Switching. A General Approach, \emph{The Journal of Derivatives},
Vol. 13, No. 1, pp. 41-47.

\item[] Elliott R. J. and Osakwe C. J. U. (2006), Option pricing for pure jump
processes with Markov switching compensators, \emph{Finance and
Stochastics}, 10, pp. 250-275.

\item[] Elliott R. J. , Siu T. K. and Chan L. (2007),
Pricing volatility swaps under Heston's stochastic volatility
model with regime switching, \emph{Applied Mathematical Finance},
Vol 14, 1, pp. 41-62.

\item[] Guo X., (2001), Information and option pricing \emph{Quantitative
Finance}, 1, pp. 38-44.

\item[] Guo X. and Zhang Q. Z (2004), Closed-form solutions
for perpetual American put options with regime switching,
\emph{SIAM Journal of Applied Mathematics}, 64 , pp. 2034-2049.

\item[] Hamilton, J.D., (1989), A new approach to the economic analysis of
non-stationary time series, \emph{Econometrica}, 57, pp. 357-384.

\item[] Hamilton, J.D., (1990), Analysis of time series subject to changes
in regime, \emph{Journal of Econometrics}, 45, pp. 39-70.

\item[] Hardy, M.R., (2001), A regime switching model of a long term
stock-returns. \emph{North American Actuarial Journal}, 3, pp.
185-211.

\item[] Heston, S. L. A. (1993), Closed-Form Solution for Options
with Stochastic Volatility with Applications to Bond and Currency
Options, \emph{Rev. Fin. Studies},  vol. 6, 327-343.

\item[] Higham N. J., (2009), The scaling and Squaring
Method for the Matrix Exponential Revisited, \emph{SIAM Review},
Vol. 51, No. 4, pp. 747-764.

\item[] F. Hubalek, J. Kallsen, and L. Krawczyk, Variance-optimal
hedging for processes with stationary independent increments,
\emph{Ann. Appl. Probab.}, 16 (2006), pp. 853-885.

\item[] T. R. Hurd and Z. Zhou, A Fourier transform method for spread
option pricing, \emph{SIAM J. Financial Math.}, 1 (2010), pp.
142-157.

\item[] Jiang Z., Pistorius M. R. On perpetual American
put valuation and first-passage in a regime-switching model with
jumps. \emph{Finance Stoch.} 12 (2008), no. 3, 331-355.

\item[] Jobert A. and Rogers L. C. G. (2006), Option pricing with
Markov-Modulated dynamics, \emph{SIAM Journal of Control and
Optimization},  Vol. 44, No. 6, pp. 2063-2078.

\item[] Khaliq, A. Q. M.; Liu, R. H. New numerical scheme for pricing
American option with regime-switching. \emph{Int. J. Theor. Appl.
Finance} 12 (2009), no. 3, 319-340.

\item[] Konikov M. and Madan D. B. (2002), Option Pricing Using
Variance Gamma Markov Chains, \emph{Review of Derivatives
Research}, Vol. 5, No. 1, pp. 81-115

\item[] Kruse S. and N\"{o}gel U. (2005), On the pricing of
forward starting options in Heston's model on stochastic
volatility, \emph{Finance and Stochastics}, 9, pp. 233-250.

\item[] Lee R. W. (2004), Option pricing by transform methods:
extensions, unifications and error control, \emph{Journal of
Computational Finance}, 7, pp. 51-86.

\item[] Lewis A.L. (2002), A simple option formula for general
jump-diffusion and other exponential L\'evy processes, Working
Paper, Optioncitynet.net.

\item[] Liu, R. H. Regime-switching recombining tree for option pricing.
\emph{Int. J. Theor. Appl. Finance} 13 (2010), no. 3, 479-499.

\item[] Liu R. H. , Zhang Q. , and Yin G. (2006), Option pricing in a regime-switching
model using the fast Fourier transform. \emph{J. Appl. Math.
Stoch. Anal.}, Vol. 22, pp.1-22.

\item[] R. Lord (2008), Efficient pricing algorithms for exotic derivatives,
PhD thesis, Univ. Rotterdam.

\item[] Naik, V. (1993), Option valuation and hedging strategies
with jumps in the volatility of asset returns. \emph{Journal of
Finance}, 48, pp. 1969-1984.

\item[] S. Raible (2000), L´evy processes in finance: Theory, numerics, and
empirical facts., PhD thesis, Univ. Freiburg.

\item[] Ramponi A. (2009), Mixture Dynamics and Regime Switching Diffusions
with Application to Option Pricing. \emph{Methodol. Comput. Appl.
Probab.}, DOI 10.1007/s11009-009-9155-1

\item[] Runggaldier  W.J. (2003) . Jump-Diffusion models, in
"Handbook of Heavy Tailed Distributions in Finance" (S.T. Rachev,
ed.), Handbooks in Finance, Book 1 (W.Ziemba Series Ed.),
Elesevier/North-Holland 2003, pp. 169-209.

\item[] Timmermann A. (2000), Moments of Markov switching
models, \emph{Journal of Econometrics}, 96, pp. 75-111.

\item[] Yao D. D., Zhang Q.  and Zhou X. Y. (2006), A
regime-switching model for European options, \emph{Stochastic
Processes, Optimization, and Control Theory Applications in
Financial Engineering, Queueing Networks, and Manufacturing
Systems} (H. M. Yan, G. Yin, and Q. Zhang, eds.), Springer, New
York, pp. 281-300.





\end{description}
\end{document}